\documentclass[fleqn,usenatbib]{mnras}

\DeclareRobustCommand{\VAN}[3]{#2}
\let\VANthebibliography\thebibliography
\def\thebibliography{\DeclareRobustCommand{\VAN}[3]{##3}\VANthebibliography}

\usepackage{newtxtext,newtxmath}
\usepackage[T1]{fontenc}
\usepackage{orcidlink}
\usepackage{tablefootnote}
\usepackage{soul}
\usepackage{comment}
\usepackage{graphicx}
\usepackage[flushleft]{threeparttable}
\usepackage{ulem}
\usepackage{xcolor}
\usepackage{subfigure}
\usepackage{multirow}
\usepackage{tabularx}
\usepackage{array}

\newcolumntype{H}{>{\setbox0=\hbox\bgroup}c<{\egroup}@{}}
\newcommand{\msolar}{M$_{\odot}$}
\newcommand{\ml}{M$_{\odot}$~yr$^{-1}$}

\newcommand{\um}{$\mu$m}

\newcommand{\Msun}{M$_{\odot}$}
\newcommand{\lsolar}{L$_{\odot}$}
\UseRawInputEncoding

\title[Unveiling Dust Reservoirs in SNe IIP]{JWST Discovery of Dust Reservoirs in Nearby Type IIP Supernovae 2004et and 2017eaw}

\author[Shahbandeh et al.]{Melissa Shahbandeh\orcidlink{0000-0002-9301-5302},$^{1,2}$\thanks{E-mail: melissa.shahbandeh@gmail.com}
Arkaprabha Sarangi\orcidlink{0000-0002-9820-679X},$^{3}$
Tea Temim\orcidlink{0000-0001-7380-3144},$^{4}$
Tam\'as Szalai\orcidlink{0000-0003-4610-1117},$^{5,6}$
Ori D. Fox\orcidlink{0000-0003-2238-1572},$^{2}$
\newauthor Samaporn Tinyanont\orcidlink{0000-0002-1481-4676},$^{7}$
Eli Dwek,$^{8}$ 
Luc Dessart,$^{9}$
Alexei V. Filippenko\orcidlink{0000-0003-3460-0103},$^{10}$
Thomas G. Brink\orcidlink{0000-0001-5955-2502},$^{10}$
\newauthor Ryan J. Foley,$^{7}$
Jacob Jencson,$^{1}$
Justin Pierel,$^{2}$ 
Szanna Zs\'iros,$^{5}$
Armin Rest,$^{2}$ 
WeiKang Zheng,$^{10}$
\newauthor Jennifer Andrews,$^{11}$ 
Geoffrey C. Clayton,$^{12}$ 
Kishalay De,$^{13}$ 
Michael Engesser,$^{2}$
Suvi Gezari,$^{2}$ 
\newauthor Sebastian Gomez\orcidlink{0000-0001-6395-6702},$^{2}$ 
Shireen Gonzaga,$^{2}$ 
Joel Johansson,$^{14}$ 
Mansi Kasliwal,$^{15}$
Ryan Lau,$^{16}$ 
Ilse De Looze,$^{17}$ 
\newauthor Anthony Marston,$^{18}$ 
Dan Milisavljevic\orcidlink{0000-0002-0763-3885},$^{19,20}$
Richard O'Steen\orcidlink{0000-0002-2432-8946},$^{2}$
Matthew Siebert,$^{2}$ 
Michael Skrutskie,$^{21}$ 
\newauthor Nathan Smith\orcidlink{0000-0001-5510-2424},$^{22}$ 
Lou Strolger,$^{2}$ 
Schuyler D.~Van Dyk\orcidlink{0000-0001-9038-9950},$^{23}$
Qinan Wang,$^{1}$
Brian Williams,$^{8}$ 
Robert Williams,$^{2}$ 
\newauthor Lin Xiao$^{24,25}$
\vspace{0.5cm}\\
$^{1}$Department of Physics and Astronomy, Johns Hopkins University, Baltimore, MD 21218, USA\\
$^{2}$Space Telescope Science Institute, 3700 San Martin Drive, Baltimore, MD 21218, USA\\
$^{3}$DARK, Niels Bohr Institute, University of Copenhagen, Jagtvej 128, 2200 Copenhagen, Denmark\\
$^{4}$Department of Astrophysical Sciences, Princeton University, Princeton, NJ 08544, USA\\
$^{5}$Department of Experimental Physics, Institute of Physics, University of Szeged, H-6720 Szeged, D{\'o}m t{\'e}r 9, Hungary\\
$^{6}$ELKH-SZTE Stellar Astrophysics Research Group, H-6500 Baja, Szegedi {\'u}t, Kt. 766, Hungary\\
$^{7}$Department of Astronomy and Astrophysics, University of California, Santa Cruz, CA 95064, USA\\
$^{8}$Observational Cosmology Lab, NASA Goddard Space Flight Center, Code 665, Greenbelt, MD 20771, USA\\
$^{9}$Institut d'Astrophysique de Paris, CNRS-Sorbonne Universit\'e, 98 bis boulevard Arago, F-75014 Paris, France\\
$^{10}$Department of Astronomy, University of California, Berkeley, CA 94720-3411, USA\\
$^{11}$Gemini Observatory, 670 N. Aohoku Place, Hilo, Hawaii, 96720, USA\\
$^{12}$Department of Physics \& Astronomy, Louisiana State University, Baton Rouge, LA 70803, USA\\
$^{13}$MIT-Kavli Institute for Astrophysics and Space Research, 77 Massachusetts Ave., Cambridge, MA 02139, USA\\
$^{14}$Department of Physics, The Oskar Klein Center, Stockholm University, AlbaNova, 10691 Stockholm, Sweden\\
$^{15}$Cahill Center for Astrophysics, California Institute of Technology, 1200 E. California Blvd. Pasadena, CA 91125, USA\\
$^{16}$NSF's NOIRLab, 950 N. Cherry Avenue, Tucson, 85719, AZ, USA\\
$^{17}$Sterrenkundig Observatorium, Ghent University, Krijgslaan 281 - S9, 9000 Gent, Belgium\\
$^{18}$European Space Agency (ESA), ESAC, 28692 Villanueva de la Canada, Madrid, Spain\\
$^{19}$Purdue University, Department of Physics and Astronomy, 525 Northwestern Ave, West Lafayette, IN 4790720, USA\\
$^{20}$Integrative Data Science Initiative, Purdue University, West Lafayette, IN 47907, USA\\
$^{21}$Department of Astronomy, University of Virginia, Charlottesville, VA 22904-4325, USA\\
$^{22}$Steward Observatory, University of Arizona, 933 N. Cherry St, Tucson, AZ 85721, USA.\\
$^{23}$Caltech/IPAC, Mailcode 100-22, Pasadena, CA 91125, USA\\
$^{24}$Department of Physics, College of Physical Sciences and Technology, Hebei University, Baoding 071002, China\\
$^{25}$Key Laboratory of High-precision Computation and Application of Quantum Field Theory of Hebei Province, Hebei University, Baoding 071002, China
}

\begin{document}
\label{firstpage}
\pagerange{\pageref{firstpage}--\pageref{lastpage}}\pubyear{2023}
\maketitle

\begin{abstract}
Supernova (SN) explosions have been sought for decades as a possible source of dust in the Universe, providing the seeds of galaxies, stars, and planetary systems. SN~1987A offers one of the most promising examples of significant SN dust formation, but until the {\it James Webb Space Telescope (JWST)}, instruments have traditionally lacked the sensitivity at both late times ($>1$~yr post-explosion) and longer wavelengths (i.e., $>10$~\um) to detect analogous dust reservoirs. Here we present {\it JWST}/MIRI observations of two historic Type IIP SNe, 2004et and SN 2017eaw, at nearly 18~yr and 5~yr post-explosion, respectively. We fit the spectral energy distributions as functions of dust mass and temperature, from which we are able to constrain the dust geometry, origin, and heating mechanism. We place a 90\% confidence lower limit on the dust masses for SNe 2004et and 2017eaw of $>0.014$ and $>4 \times 10^{-4}$ \Msun, respectively. More dust may exist at even colder temperatures or may be obscured by high optical depths. We conclude dust formation in the ejecta to be the most plausible and consistent scenario. The observed dust is radiatively heated to $\sim 100$--150~K by ongoing shock interaction with the circumstellar medium. Regardless of the best fit or heating mechanism adopted, the inferred dust mass for SN 2004et is the second highest (next to SN 1987A) inferred dust mass in extragalactic SNe thus far, promoting the prospect of SNe as potential significant sources of dust in the Universe.
\end{abstract}


\section{Introduction}\label{sec:intro}
The source of the large amounts of dust observed in high-redshift galaxies remains uncertain \citep{Maiolino_2004}.
For over 50~yr, core-collapse supernovae (CCSNe) have been considered as possible sources of dust \citep{Cernuschi_1967, Hoyle_1970}. 
Type II-P supernovae (SNe~IIP), in particular, are likely candidates since they account for $\sim 70$\% of all SNe~II, and are plentiful enough that they could account for the observed dust at high redshifts \citep[e.g., see][and references therein]{Schneider_2004, Dwek_2007, Nozawa_2008, Gall_2011_a}. 
Models of expanding SN~IIP ejecta succeed in condensing out sufficient quantities (0.1--1~\Msun; \citealt{Todini_2001,Nozawa_2003,Cherchneff_2009, Sarangi_2015, Sluder_2018, Sarangi_2018}).

\begin{figure*}
\centering
\includegraphics[width=0.7\textwidth]{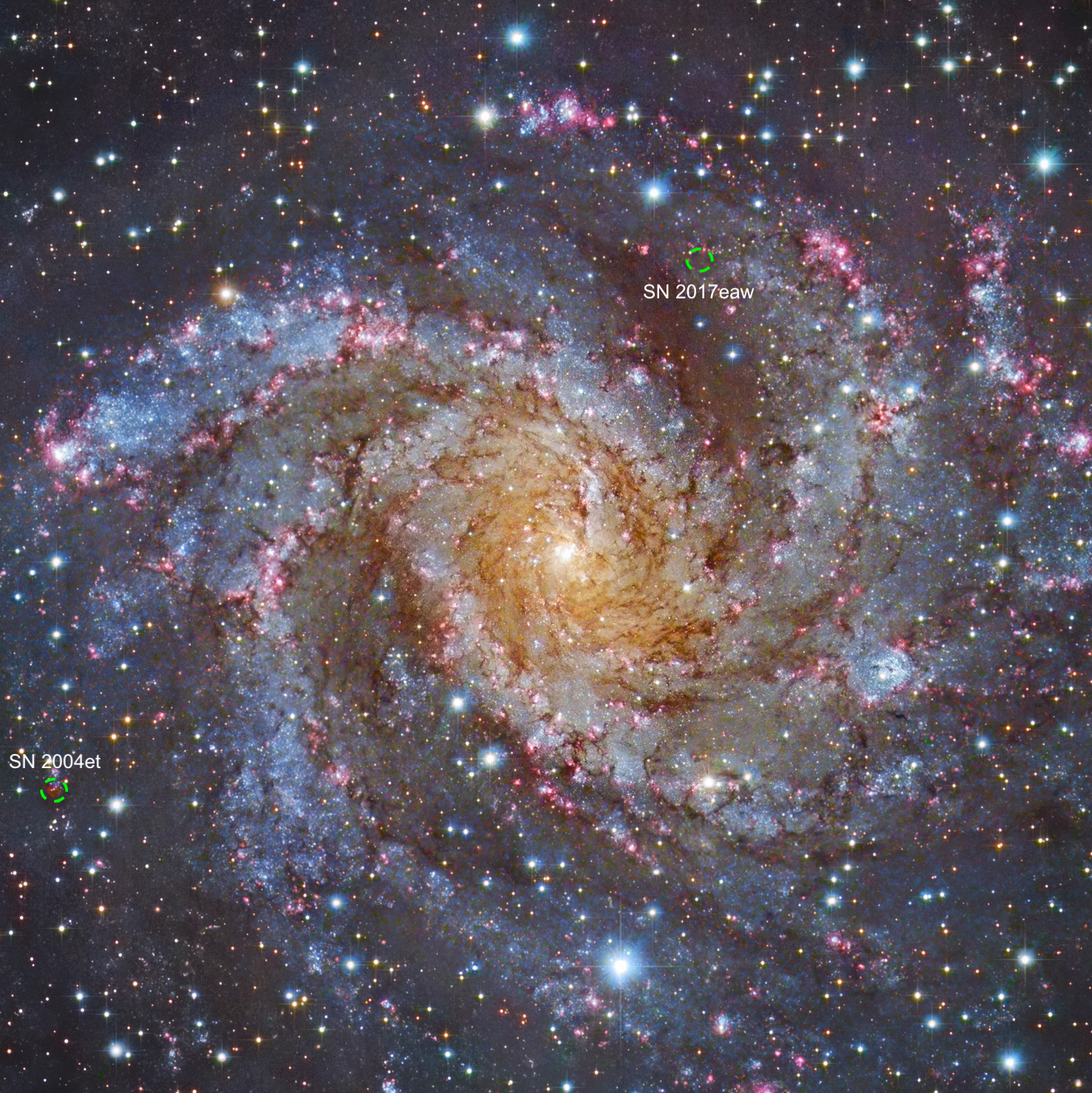}
\includegraphics[width=0.49\textwidth]{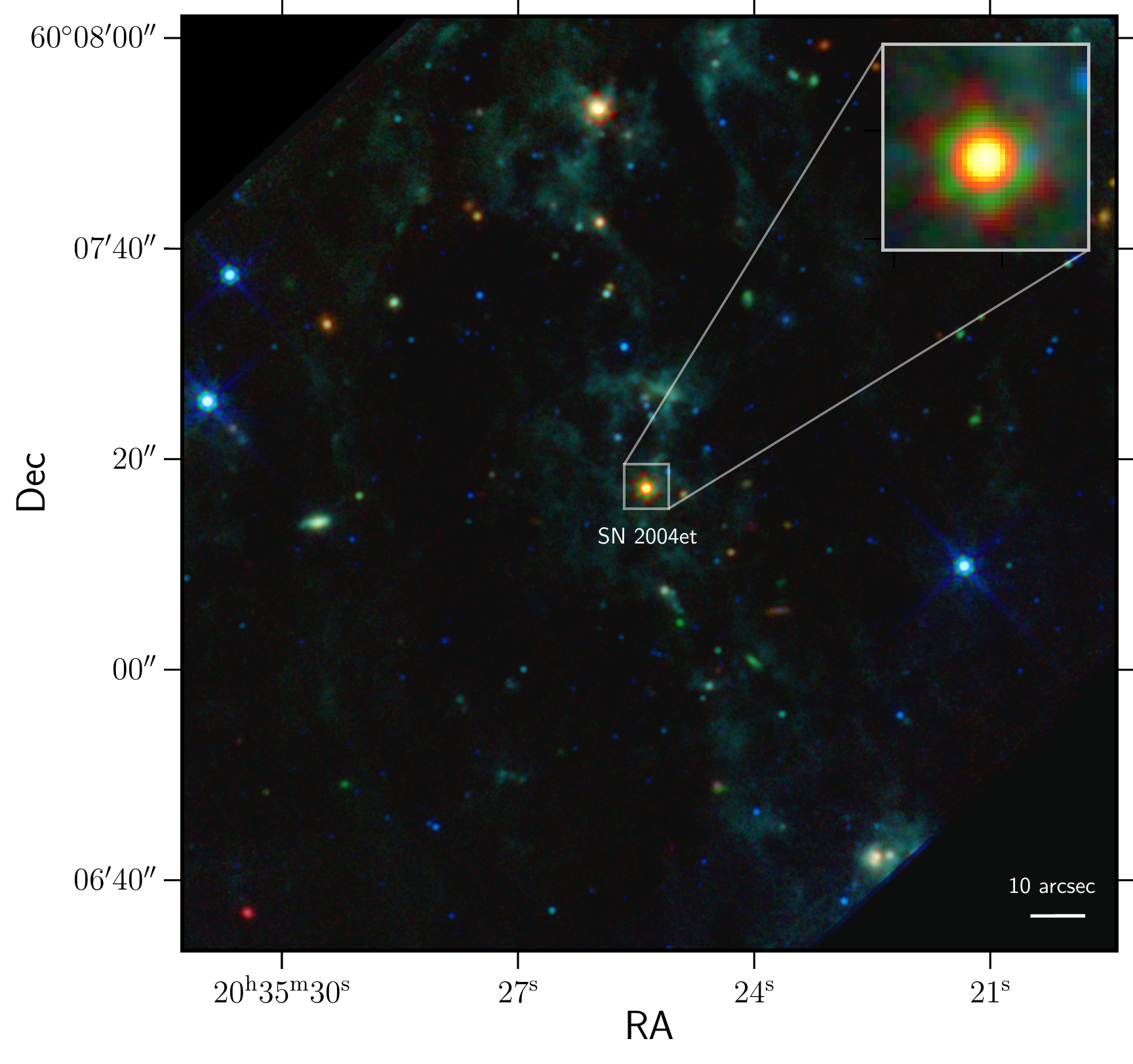}
\includegraphics[width=0.49\textwidth]{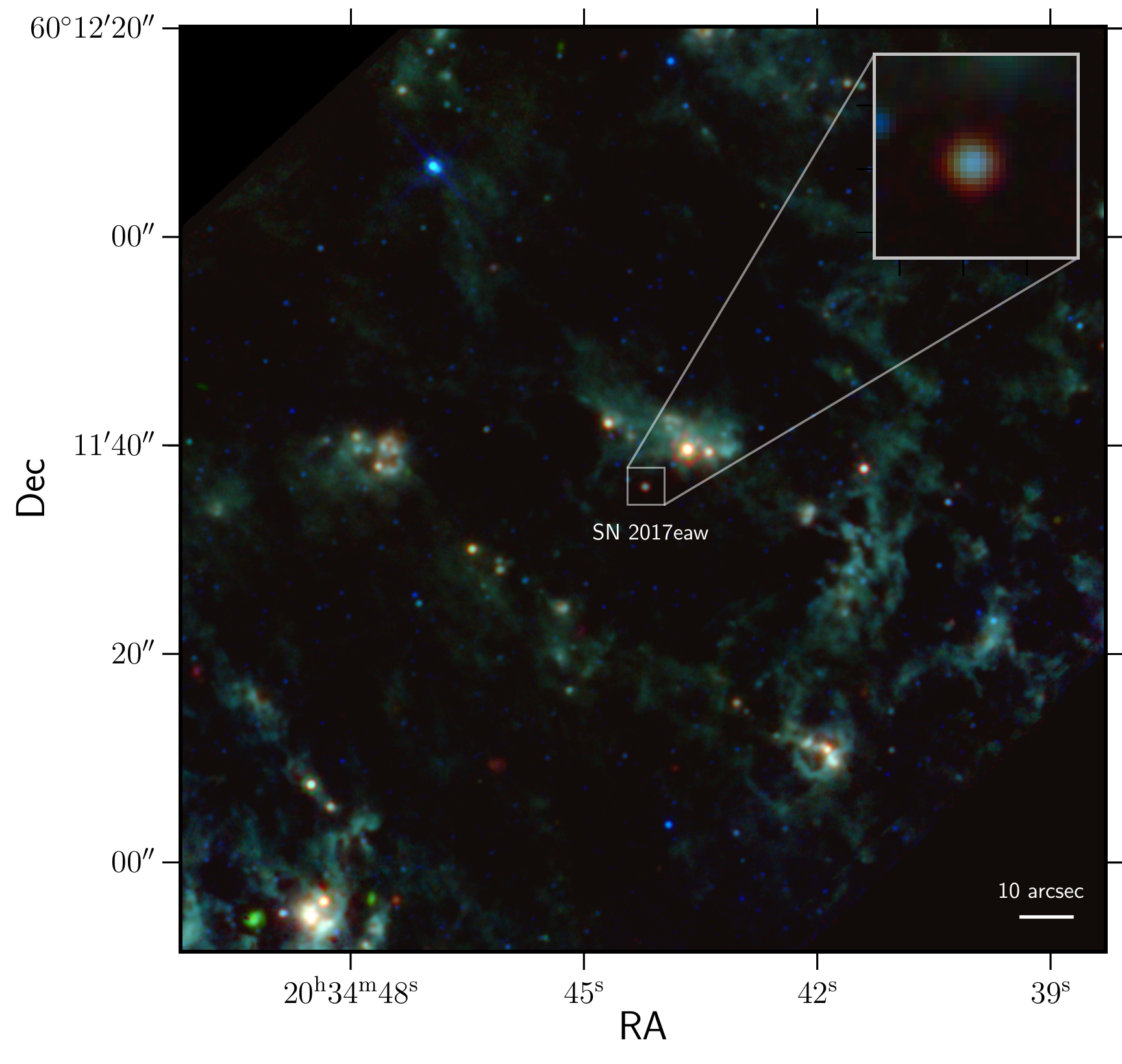}
    \caption{\textbf{Top panel:} {\it HST} image of NGC 6946, the host galaxy of both SNe 2004et and 2017eaw (courtesy of NASA, ESA, STScI, Gendler, and the Subaru Telescope). \textbf{Bottom-left panel:} Composite image of SN~2004et taken with {\it JWST}'s MIRI. \textbf{Bottom-right panel:} Composite image of SN 2017eaw taken with {\it JWST}'s MIRI. (MIRI filters used: F560W, F1000W, F1130W, F1280, F1500W, F1800W, F2100W, F2550W.)}
    \label{fig:host}
\end{figure*}

A number of SNe of different subclasses were observed by the {\it Spitzer Space Telescope} (hereafter {\it Spitzer}), but measured dust masses were often too small to explain the amount of dust at high redshifts. 
Only a handful of nearby SNe~II, let alone SNe~IIP, have shown direct observational evidence for dust condensation, and these examples all yielded 2--3 orders of magnitude ($<10^{-2}$~\Msun) less dust than predicted by the above models \citep[e.g.,][]{Gerardy_2002, Pozzo_2004, Sugerman_2006, Meikle_2007, Meikle_2011, Andrews_2010, Andrews_2011, Fabbri_2011, Szalai_2011}. 
Renewed interest in SN dust began with several important advances: (i) the far-infrared/submm/mm detection of a large amount of cold dust in SN~1987A \citep{Matsuura_2011, Indebetouw_2014,bevan16}, (ii) the detection of large quantities of dust in some Galactic SNRs \citep[e.g.,][]{Owen_2015,Temim_2017,delooze17,priestly20}, (iii) large dust masses ($>10^{-2}$~\msolar) inferred from optical line profile asymmetries \citep{bevan16,bevan17,bevan19} and (iii) the identification of isotopic anomalies in meteorites \citep[e.g.,][]{Clayton_2004}. 

Despite these promising discoveries, large dust reservoirs have not yet been directly observed at mid-IR wavelengths in extragalactic SNe, which could be due to a number of reasons. 
New dust condenses at high temperatures, but its temperature decreases rapidly with the expansion of the ejecta and the decay of internal radioactive heating sources. Consequently, the reservoirs of warm and cold (i.e., not hot) dust may only become detectable at longer wavelengths \citep{Wesson_2015}. 
{\it Spitzer} was limited in its sensitivity at these wavelengths during the Cold Mission, which ended in 2008. 
After that point, {\it Spitzer} had access to only 3.6 and 4.5~\um\ imaging. 
Furthermore, the timeline of dust growth is quite uncertain. 
\citet{Gall_2014} suggest that accelerated dust growth does not begin until $>5$~yr after explosion. 
However, \citet{Dwek_2019} argue that the dust mostly forms at earlier epochs ($<2$~yr) but is largely obscured due to optical-depth effects until the ejecta expand sufficiently and become optically thin at later epochs. 
The {\it Spitzer} observations could probe only the warmer dust components ($>500$~K) and/or earlier epochs ($<5$~yr). For the first time, {\it JWST's} Mid-Infrared Instrument (MIRI) offers the possibility to probe cooler dust growth in extragalactic SNe \citep[e.g.,][]{Hosseinzadeh_2022}. 

Here we present MIRI imaging observations of two SNe~IIP, 2004et and 2017eaw, as part of {\it JWST} GO-2666 (PI O. Fox).
SN~2004et was discovered by S. Moretti on 2004 September 27 UT \citep{Zwitter_2004} and SN 2017eaw was discovered on 2017 May 14.24 UT \citep{Wiggins_2017,Dong_2017}. Both SNe located in the nearby, face-on spiral galaxy NGC~6946, which has produced around a dozen known SNe and other luminous transients (including SNe~IIP 1948B, 2002hh, the Type IIL SN~1980K, and the ``SN impostor" SN~2008S). 
While they appeared on opposite sides of the host galaxy, both photometric and spectral evolution of SNe 2004et and SN~2017eaw seem to be quite similar and correspond well to those of ``normal'' SN~IIP explosions; however, SN~2004et was a bit more luminous than SN 2017eaw \citep[they reached a peak $V$ magnitude of 12.6 and 12.8, respectively; see, e.g.,][]{Sahu_2006,Maguire_2010,Tsvetkov_2018,VanDyk_2019,Szalai_2019,Buta_2019}.

The search for potential progenitors of both SNe in archival imaging started right after their discoveries. 
For SN~2004et, \citet{Li_2005} identified a yellow supergiant on Canada-France-Hawaii Telescope images. 
Later, \citet{Crockett_2011} showed that this source was still visible years after explosion and -- based on high-resolution post-explosion images obtained by William Herschel Telescope, {\it Hubble Space Telescope (HST)}, and the Gemini telescope -- the progenitor is rather a late-K to late-M supergiant of $M_{\rm prog} = 8^{+5}_{-1}$~M$_{\odot}$. 
Nevertheless, hydrodynamic and semi-analytical modeling of optical light curves lead to an ejecta mass of $\sim 11$--23~M$_{\odot}$ \citep{Utrobin_2009,Nagy_2014,Morozova_2018b,Ricks_2019,Martinez_2019}, implying a much larger progenitor mass. Using the method of late-time spectral modeling, \citet{Jerkstrand_2012} determined a value of $M_{\rm prog} \approx 15$~M$_{\odot}$ for SN~2004et.

In the case of SN~2017eaw, archival {\it HST} and Large Binocular Telescope images enabled the identification of the progenitor as a red supergiant (RSG) star with an estimated initial  mass of $M_{\rm prog} \approx 11$--17~M$_{\odot}$ \citep{VanDyk_2017,Kilpatrick_2018,Johnson_2018,Rui_2019}. Modeling of optical light curves and nebular spectra \citep{Szalai_2019}, as well as of near-infrared (NIR) spectra \citep{Rho_2018}, consistently give $M_{\rm prog} \approx 15$~M$_{\odot}$. 
Note that \citet{Williams_2018} give a much lower value for the progenitor mass (8.8$^{+2.0}_{-0.2 }~M_\odot$) of SN~2017eaw from modeling the local stellar population.

Both SNe also show evidence for early dust formation in their ejecta. For SN~2004et, \citet{Kotak_2009} and \citet{Fabbri_2011} present one of the most detailed mid-IR {\it Spitzer} datasets of an SN including four-channel IRAC (3.6--8.0~\um), MIPS (24.0~\um), and IRS measurements. For SN~2017eaw, near-IR spectra and photometry, accompanied by 3.6~\um\ and 4.5~\um\ {\it Spitzer} photometry, were used for giving estimates of the properties of early-time dust \citep[][see details later]{Rho_2018,Tinyanont_2019}.

SNe 2004et and 2017eaw belong to the small group of SNe~IIP showing signs of late-time circumstellar interaction. Early-time radio (SN~2004et: \citealt{Stockdale_2004,Beswick_2004,Misra_2007,Marti-Vidal_2007}; SN~2017eaw: \citealt{Argo_2017,Nayana_2017}) and X-ray (SN~2004et: \citealt{Misra_2007,Rho_2007}; SN 2017eaw: \citealt{Kong_2017,Grefensetette_2017}) detections, giving the highest fluxes among SNe~II-P, have already implied the presence of (moderate) circumstellar material (CSM) in the vicinity of both SNe \citep{Chevalier_2006,Szalai_2019}. In the case of SN~2017eaw, an early bump peaking at $\sim 6$--7~days after explosion in all optical bands also supports this picture \citep{Szalai_2019}. Moreover, based on archival {\it Spitzer} images, \citet{Kilpatrick_2018} showed that the progenitor of SN~2017eaw was surrounded by a dusty shell at $\sim 4000$~R$_{\odot}$. Final evidence for the presence of CSM was the detection of strong H$\alpha$ emission with a box-like line profile in both SNe $\sim 2.5$~yr after explosion \citep{Kotak_2009,Maguire_2010,Weil_2020} and very late-time optical detections \citep{rizzo22}; these findings also strengthen the picture on the similarity of the two exploded stars and of their environments.

Combined light-curve analysis of SNe 2004et and 2017eaw also enabled a proper estimate of the distance of the host galaxy NGC~6946. While previous work gave a value of $D \approx 5.5$--5.9~Mpc, \citet{Szalai_2019} showed that the distance of NGC~6946 is probably larger by $\sim 30$\%, in agreement with the values given by the tip-of-the-red-giant-branch (TRGB) method \citep{Tikhonov_2014, Anand_2018}. In this paper, we use the latest TRGB distance\footnote{\url{http://edd.ifa.hawaii.edu/get_cmd.php?pgc=65001}} of $D = 7.12$~Mpc ($\mu = 29.26$~mag).


In Section~\ref{sec:obs} of this paper, we present our observations and reductions.  Section~\ref{sec:dustmass} describes the dust formation physical scenarios and our dust models. We interpret the results in Section~\ref{sec:discussion} and discuss the dust masses relative to those of other historic SNe. A summary of our results and the conclusions are given in Section~\ref{sec:conclusion}.
\section{Observations and Data Reduction} \label{sec:obs}
\subsection{{\sl JWST\/}/MIRI Imaging}

As part of the Cycle 1 General Observers (GO) 2666 program (PI O. Fox), we obtained images of SNe~2004et and 2017eaw with the {\it JWST} Mid-Infrared Instrument (MIRI; \citealt{Bouchet_2015, Ressler_2015, Rieke_2015, Rieke_2022}) on 20~Sep.~2022 (6567 days post-explosion) and 21~Sep.~2022 (1954 days post-explosion), respectively (see Figure~\ref{fig:host}).
The observations were acquired in the F560W, F1000W, F1130W, F1280W, F1500W, F1800W, F2100W, and F2550W filter bands, using the FASTR1 readout pattern in the FULL array mode and a 4-point extended source dither pattern. 
The data were processed with the {\it JWST} Calibration Pipeline version 1.7.2, using the Calibration Reference Data System version 11.16.9. 

A ``background image" was constructed for each filter by taking a sigma-clipped average of the individual dithers in detector coordinates, and this background was then subtracted from the calibrated (level two) individual dither images in the corresponding filter.
The background-subtracted level-two images were then combined into a single calibrated image for each filter \citep{Bright_2016, Greenfield_2016, bushouse_howard_2022}.
To measure the brightness of both SNe, we performed 
point-spread-function (PSF) photometry on background-subtracted level-two data products using WebbPSF \citep{Perrin_2014}.
In order to calibrate the flux, we applied flux offsets by measuring the PSF of all the stars in the field and comparing them to the corresponding catalogs created by the pipeline. 
The fluxes of all four dithers of each filter were then averaged.
Table \ref{tab:JWST_phot} summarizes the PSF photometry for both SNe in terms of both flux and AB magnitude \citep{Oke_1983}.

\begin{table}
\centering
\caption{{\it JWST}/MIRI Photometry\tablefootnote{All observations were obtained on MJD 59842} \label{tab:JWST_phot}}
\setlength{\tabcolsep}{3pt}
\begin{tabular}{ l c c c c}
\hline
\hline
\multicolumn{1}{l}{Filters}&
\multicolumn{2}{c}{SN~2004et}&
\multicolumn{2}{c}{SN~2017eaw}\\
&Mag &Flux&Mag &Flux\\
& &[$\mu$Jy]& &[$\mu$Jy]\\
\hline
F560W   &  20.89 $\pm$ 0.03   &    15.96 $\pm$ 0.49    &   21.90 $\pm$ 0.04    &   6.39 $\pm$ 0.29\\
F1000W   &  18.67 $\pm$ 0.01    &   122.63 $\pm$ 1.51     &   19.49 $\pm$ 0.01    &  58.12 $\pm$ 0.79 \\
F1130W   &  17.78 $\pm$ 0.01   &   279.68 $\pm$ 3.60     &   19.64 $\pm$ 0.02    &  51.17 $\pm$ 1.25  \\
F1280W   &  17.25 $\pm$ 0.01   &    456.39 $\pm$ 4.36    &   19.81 $\pm$ 0.01    &   43.30 $\pm$ 0.58 \\
F1500W   &   16.74 $\pm$ 0.01  &   727.80 $\pm$ 5.85     &   19.40 $\pm$ 0.01    &   63.26 $\pm$ 0.69 \\
F1800W   &  16.36 $\pm$ 0.01    &    1037.30 $\pm$ 6.42    &    18.73 $\pm$ 0.01    &  117.00 $\pm$ 1.28 \\
F2100W   &  16.16 $\pm$ 0.01   &    1238.45 $\pm$ 7.58    &    18.63 $\pm$ 0.01   &  128.20 $\pm$ 1.61  \\
F2550W   &   15.75 $\pm$ 0.01  &    1804.75 $\pm$ 21.60    &    18.87 $\pm$ 0.05   &  102.47 $\pm$ 6.08 \\
\hline
\end{tabular}
\end{table}

\subsection{{\it HST} Archival Photometry}

For part of this analysis (Section \ref{sec:dust_geometry} below), we obtained {\it HST} images for the two SNe from the Mikulski Archive for Space Telescopes (MAST). Although all the data were previously published in separate papers, we reanalyze the data here to provide single, consistent set of photometry reduced with the latest version of the {\it HST} pipeline.
We downloaded drizzled images from MAST, so they have been processed through the standard pipeline at the Space Telescope Science Institute (STScI). 
Drizzled images are a combination of individual bias-subtracted, dark-subtracted, flat-fielded exposures that are then corrected for geometric distortion. We used the \textit{JWST}-\textit{HST} Alignment Tool (\texttt{JHAT}\footnote{\url{readthedocs.jhat.io}}), which applies a relative astrometric correction to each image by iteratively matching sources between observations, to  ensure precise alignment between the \textit{HST} and \textit{JWST} data. 

With all images aligned, the total SN flux was measured inside a circular aperture placed in each \textit{HST} observation at the \textit{JWST} SN location using the Python package \texttt{photutils}. The aperture radius was set to 3 pixels, or $\sim 1.5$--2 times the full width at half-maximum intensity for \textit{HST} ACS and WFC3, owing to a nearby star contaminating the flux at larger radii.
We subtract the local background flux from the aperture flux with the $\sigma$-clipped (using 3$\sigma$) median value in a circular annulus centered at the SN location, using an inner radius of 5 pixels and an outer radius of 7 pixels. 
This is sufficiently large that $>99.9\%$ of the SN flux should be within the annulus inner radius, but small enough to ensure the neighboring star contamination and background light are removed. 
Finally, the measured flux for each image was corrected using aperture corrections from the public ACS/WFC3 encircled energy tables\footnote{\url{https://www.stsci.edu/hst/instrumentation/acs/data-analysis/aperture-corrections}},\footnote{\url{https://www.stsci.edu/hst/instrumentation/wfc3/data-analysis/photometric-calibration/uvis-encircled-energy}}, and converted to AB magnitudes using the time-dependent inverse sensitivity and filter pivot wavelengths provided with each data file. 

\subsection{{\it Spitzer} Archival Photometry}
\label{sec:spitzer}

SN~2004et was observed with the \textit{Spitzer} InfraRed Array Camera (IRAC; \citealp{Fazio_2004}) from 2004 to 2019.
Data during the Cold Mission of \textit{Spitzer}, prior to May 2009, were obtained in all four IRAC channels, while later data during the Warm Mission were obtained in the 3.6~\um\ and 4.5~\um\ channels. 
Photometry up to 1803 days post-explosion was published by \cite{Kotak_2009, Fabbri_2011}.
The bulk of the late-time data after 2014 was obtained as part of the SPitzer InfraRed Intensive Transients Survey (SPIRITS; \citealp{Tinyanont_2016, Kasliwal_2017, Jencson_2019}). 

To perform new photometry on the 3.6~\um\ and 4.5~\um\ IRAC data, we used a single archival pre-explosion \textit{Spitzer} image 
to estimate and remove the galaxy background and nearby source contamination.
We rotated and aligned IRAC images containing SN light based on the sky coordinates supplied in \textit{Spitzer} data, and then simply subtract the archival image from a new science image. 
The location around SN~2004et is sufficiently sparse that careful PSF matching is not crucial. 
We conducted aperture photometry on the background-subtracted images and applied appropriate aperture corrections as given by the IRAC instrument handbook. 
We also estimated and removed residual background in the subtracted images using a sky aperture offset from the SN. 

\begin{figure*}
    \centering
    \includegraphics[width=1\textwidth]{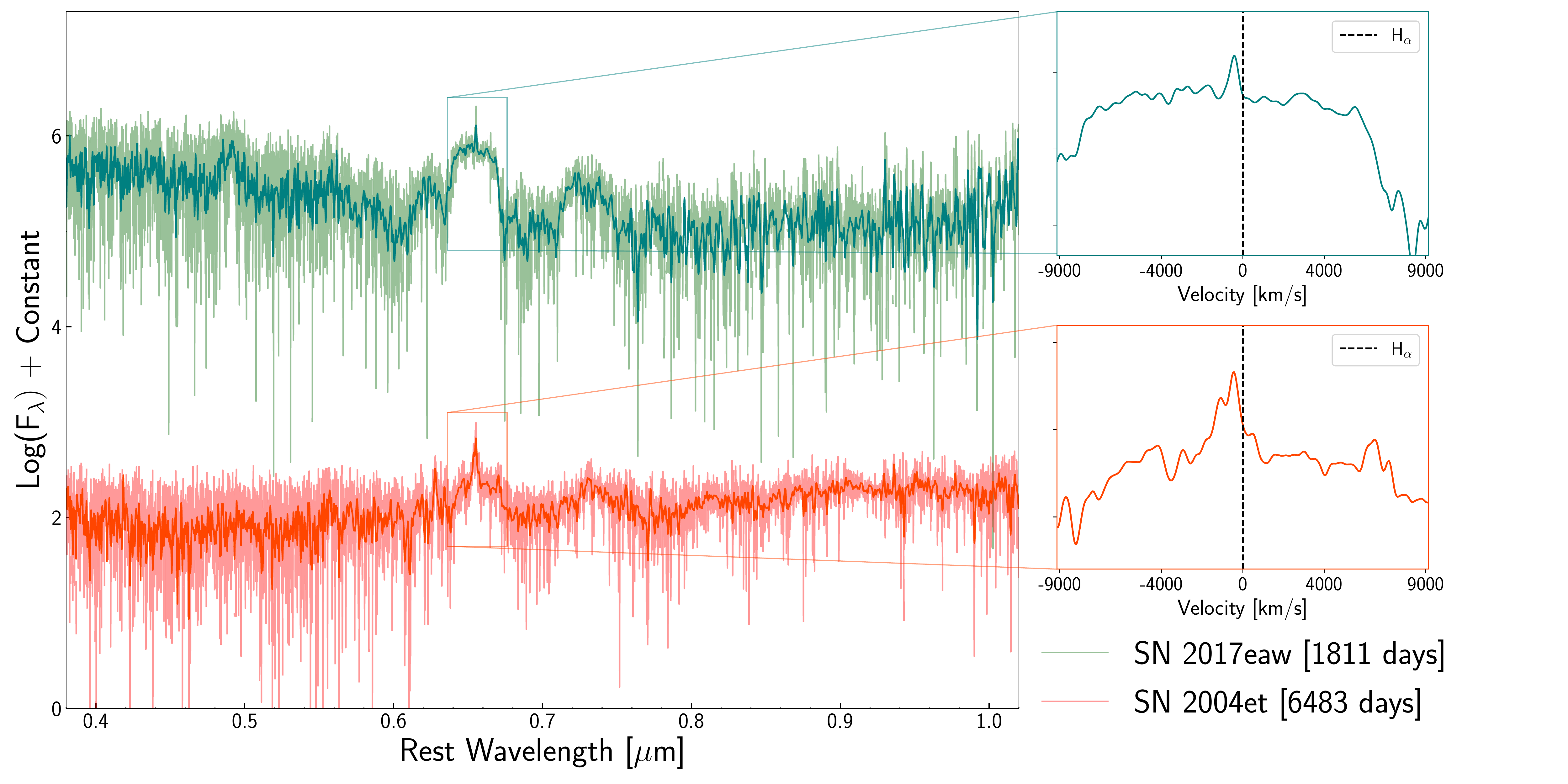}\hfill
    \caption{Late-time spectra of SNe 2004et and 2017eaw were obtained with the Keck Low Resolution Imaging Spectrometer (LRIS) on days 6424 and 1811. The presence of the H$\alpha$~line signifies ongoing shock interaction with a pre-existing circumstellar medium (CSM).}
    \label{fig:opticalspec}
\end{figure*}

\subsection{{\it WISE} Archival Photometry of SN 2004et}
\label{sec:wise}

The location of SN~2004et
was observed at several epochs during the ongoing NEOWISE all-sky, mid-IR survey in the $W1$ ($3.4$~\um) and $W2$ ($4.6$~\um) bands \citep{Wright_2010, Mainzer_2014}. We obtained the coadded time-series images of the fields created as part of the unWISE project \citep{Lang_2014, Meisner_2018}. A custom code \citep{De_2020} based on the ZOGY algorithm \citep{Zackay_2016} was used to perform image subtraction on the NEOWISE images using the full-depth coadds of the {\it WISE} and {\it NEOWISE} missions (obtained during 2010--2014) as reference images. Photometric measurements were obtained by performing forced PSF photometry at the transient position on the difference images until the epoch of the last {\it NEOWISE} data release (data acquired until December 2021).

By mid-2017 ($\sim 4500$~days post-discovery), SN~2004et had faded to a consistent flux below the zero-level of $16 \pm 9$ and $76 \pm 30$~$\mu$Jy in the W1 and W2 difference images, respectively. This implies that SN emission was present in the 2010--2014 reference image coadd, and is largely consistent with the fluxes measured by \textit{Spitzer} at those epochs in the similar 3.6~\um\ and 4.5~\um\ IRAC bands. We thus offset the difference measurements by these values to obtain absolute flux estimates, though this yields no positive detections of the SN ($>3\sigma$) after 2014. 

\subsection{Optical Spectroscopy}

Late-time spectra of SNe 2004et and 2017eaw were obtained with the Keck Low Resolution Imaging Spectrometer (LRIS; \citealp{Oke_1995}) on days 6424 and 1811, respectively (Figure \ref{fig:opticalspec}). The spectra were acquired with the slit oriented at or near the parallactic angle to minimize slit losses caused by atmospheric dispersion \citep{Filippenko_1982}. The LRIS observations utilized the $1\arcsec$ slit, 600/4000 grism, and 400/8500 grating, to produce a similar wavelength range and spectral resolving power. Data reduction followed standard techniques for CCD processing and spectrum extraction \citep{Silverman_2012}
utilizing IRAF\footnote{IRAF is written and supported by the National Optical Astronomy Observatories, operated by the Association of Universities for Research in Astronomy, Inc. under cooperative agreement with the National Science Foundation.} \citep{Tody_1986} routines and custom Python and IDL codes.\footnote{\url{https://github.com/ishivvers/TheKastShiv}}
The most recent LRIS spectra (from 2017 and later) were processed using the LPipe data-reduction pipeline \citep{Perley_2019}. Low-order polynomial fits to comparison-lamp spectra were used to calibrate the wavelength scale, and small adjustments derived from night-sky lines in the target frames were applied. The spectra were flux calibrated using observations of appropriate spectrophotometric standard stars observed on the same night, at similar airmasses, and with an identical instrument configuration.

\section{Analysis} \label{sec:dustmass}

\subsection{Physical Scenarios}
\label{sec:scenarios}

\begin{figure}
    \centering        
    \includegraphics[width=0.45\textwidth]{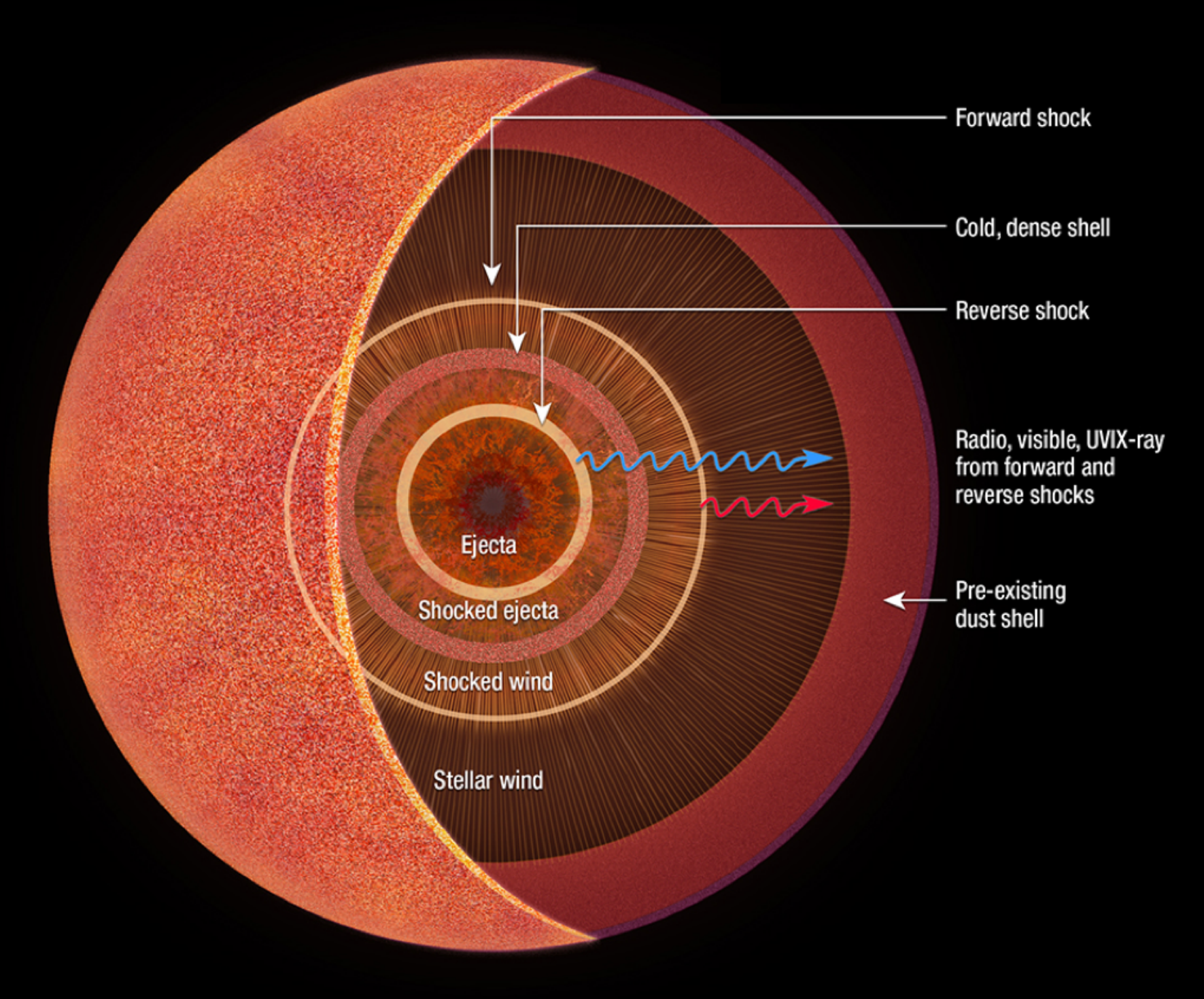}
        \caption{Illustration of how the dusty SN environment is relevant to our understanding of the SN explosion and progenitor taken from a number of papers \citep[e.g.,][and those within]{Fox_2010}. The dust may be newly formed in the ejecta or a cool, dense shell behind the forward shock. It may also be pre-existing as a circumstellar shell formed by pre-SN mass-loss from the progenitor star. The dust may be directly heated or radiatively heated by optical, UV, and X-ray emission from the shock. Disentangling all of these possibilities can constrain the origin of dust in the Universe and the late-stages of massive star evolution.}
    \label{fig:illustration}
\end{figure}

Late-time thermal-IR continuum emission typically indicates the presence of dust. 
The general SN environment can be complicated, with many different possible origins and heating mechanisms for the dust (Figure~\ref{fig:illustration}). 
The dust may be newly formed or pre-existing at the time of the SN.  If newly formed, the dust may condense in the expanding SN ejecta \citep{Kozasa_1989, Wooden_1993} or in the cool, dense shell of post-shocked gas lying in between the forward and reverse shocks \citep{Smith_2008b}. If pre-existing, the dust may have formed in a steady wind or during a short duration pre-SN outburst. In any scenario, several heating mechanisms are possible, including a thermal light echo from the peak SN flash, radiative heating from circumstellar interaction, and/or collisional heating by hot gas in the reverse shock \citep[e.g.,][]{Fox_2010}. While strong CSM interaction that yields a Type IIn event is not typically associated with SNe~IIP given the low density, relatively low mass-loss rates of their progenitor stars' winds ($10^{-6}$~\ml; \citealt{Beasor_2020}), it was more recently showed that even modest winds can build up and result in a relatively large ultraviolet (UV) flux at late times \citep{Dessart_2022,dessart23}.

\subsection{Corresponding Dust Models}
\label{sec:dustmodels}


In the optically thick case, the actual dust mass $M_{\rm dust}$ is related to the observationally inferred dust mass, $M_{\rm dust}^{\rm obs}$,  by  \citep{Fox_2010, Dwek_2019}

\begin{equation}\label{eqn:mass} 
M_{\rm dust} = {M_{\rm dust}^{\rm obs} \over P_{\rm esc}(\tau)},
\end{equation}

where
\begin{equation}\label{eqn:thick} 
M_{\rm dust}^{\rm obs} = {F_{\lambda}^{\rm obs}(\lambda)\ d^2\ \over B_{\lambda}(\lambda, T_{\rm dust})\, \kappa(\lambda)}
\end{equation}

and $F_{\lambda}^{\rm obs}(\lambda)$ is the specific flux, $d$ the distance to the source, 
 $\kappa(\lambda)$ is the dust mass absorption coefficient, $B_{\lambda}(\lambda, T_{\rm dust})$ is the Planck function at wavelength $\lambda$, $T_{\rm dust}$ is the dust temperature, and $P_{\rm esc}$ is the escape probability of the infrared photons from the emitting region.

As described by \citet{Cox_1969} and \citet{Osterbrock_2006}, $P_{\rm esc}$ is a function of optical depth ($\tau$). For a homogeneous dusty sphere in which the absorbers (dust) are uniformly mixed with the emitting sources (in this case dust as well) is given by 
\footnote{The $\lambda$ and time dependence of $\tau$ has been suppressed for the sake of clarity.}

\begin{equation}\label{eqn:Pesc}
P_{\rm esc}(\tau) = {3\over 4 \tau}\ \left[1-{1\over2\tau^2}+({1\over\tau}+{1\over2\tau^2})e^{-2\tau}\right] \, .
\end{equation}

The parameter $\tau$ is a typical optical depth, which depends on the geometry of the emitting dust. Assuming that the dust in the SN ejecta can be approximated by a homogeneously expanding sphere, we get that

\begin{eqnarray}\label{eqn:tau}
\tau(\lambda, t) & = & \rho_\rmn{dust}(t)\, R(t)\, \kappa(\lambda) = {3\over 4}\, \left({M_\rmn{dust}(t) \over \pi R(t)^2}\right) \kappa(\lambda) \nonumber \\
 & = & {3\over 4}\, \left({M_\rmn{dust}(t)\over \pi v_\rmn{ej}^2}\right) \kappa(\lambda)\, t^{-2} 
\end{eqnarray}

where $\rho_{\rm dust}(t)$~is the ejecta density, $M_{\rm dust}(t)$ is the total mass of ejecta dust, $R(t)$ is the ejecta radius at time $t$, and $v_\rmn{ej}$ is the ejecta velocity.

As a reference, 1~$M_{\odot}$ of ejecta dust expanding at a velocity of 5000~km~s$^{-1}$ will have a radial optical depth of $\approx$ 6 at 20~$\mu$m after 10 years of expansion, where we adopted a value of $\kappa \approx 300$~cm$^2$~g$^{-1}$, which is applicable for both silicates or amorphous carbon dust. So a priori, we cannot rule out the possibility that a large amount of dust may be hidden because of optical depth effects.

We can also calculate a {\it maximum} optical depth by assuming that the radiating dust is homogeneously distributed over the minimally allowed radius, given by the radius, $R_{BB},$ derived by assuming that the observed flux emanates from a perfect blackbody

\begin{equation}
\label{equ:rad}  
R_{\rm BB}(t)=\sqrt{\frac{L}{4 \pi \sigma T^4}}=\sqrt{\frac{4 \pi d^2 \int F_\lambda^{\text {obs }} d \lambda}{4 \pi \sigma T^4}}
\end{equation}

For values of $R_{BB} \approx 5\times 10^{16}$~cm, we get a maximum 20 $\mu$m optical depth of 60.

\begin{table*}
\centering
\caption{Best-Fitting Dust Model Parameters for SN~2004et
\label{tab:model_fits04et}}
\setlength{\tabcolsep}{10.pt}
\renewcommand{\arraystretch}{1.5}
\begin{tabular}{lccccc}
\hline\hline
Model   &   $M_{\rm dust}$$^{\rm a}$  &   T$^{\rm a}$   &   R$^{\rm a}$   &   $\tau^{\rm b}$  &   $P_{\rm esc}^{\rm b}$   \\
    &   [\Msun] &   [K] &   [cm]    &   &   \\
\hline
(a) Dusty Sphere (Amorphous C) &  0.044$^{+...}_{-0.007}$&   146$^{+20}_{-10}$     &   1.19$^{+...}_{-0.63} \times10^{17}$   & 0 $< \tau <$ 4.1  &  0.17 $< P_{\rm esc} <$ 1 \\
(b) Optically Thin Dust (Amorphous C) &  0.047$^{+0.007}_{-0.006}$&   137$^{+3}_{-3}$     &   --   & $<<1$  &  $\approx1$ \\
(c) Dusty Sphere (Silicates) &  0.036$^{+...}_{-0.012}$&   167$^{+5}_{-5}$     &   5.24$^{+0.46}_{-0.42} \times10^{16}$   & 8.9 $< \tau <$ 29.7  &  0.02 $< P_{\rm esc} <$ 0.08 \\
(d) Optically Thin Dust (Silicates) &  0.012$^{+0.008}_{-0.005}$&   136$^{+10}_{-10}$     &  --    & $<< 1$  &  $\approx1$\\
\end{tabular}
\begin{tablenotes}
      \small
      \item $^{\rm a}$ The numbers reported here are the best fit $\pm$~the 1$\sigma$ (68\%) confidence interval as reported by \texttt{lmfit.conf\_interval}. See Table \ref{tab:final_properties} for final reported numbers.\\
      \item $^{\rm b}$ $\tau$ and $P_{\rm esc}$ values have been calculated at the largest flux value ($\lambda \approx$ 16~\um).\\
    \end{tablenotes}
\end{table*}

\begin{figure*}
    \centering
    \includegraphics[width=0.5\textwidth]{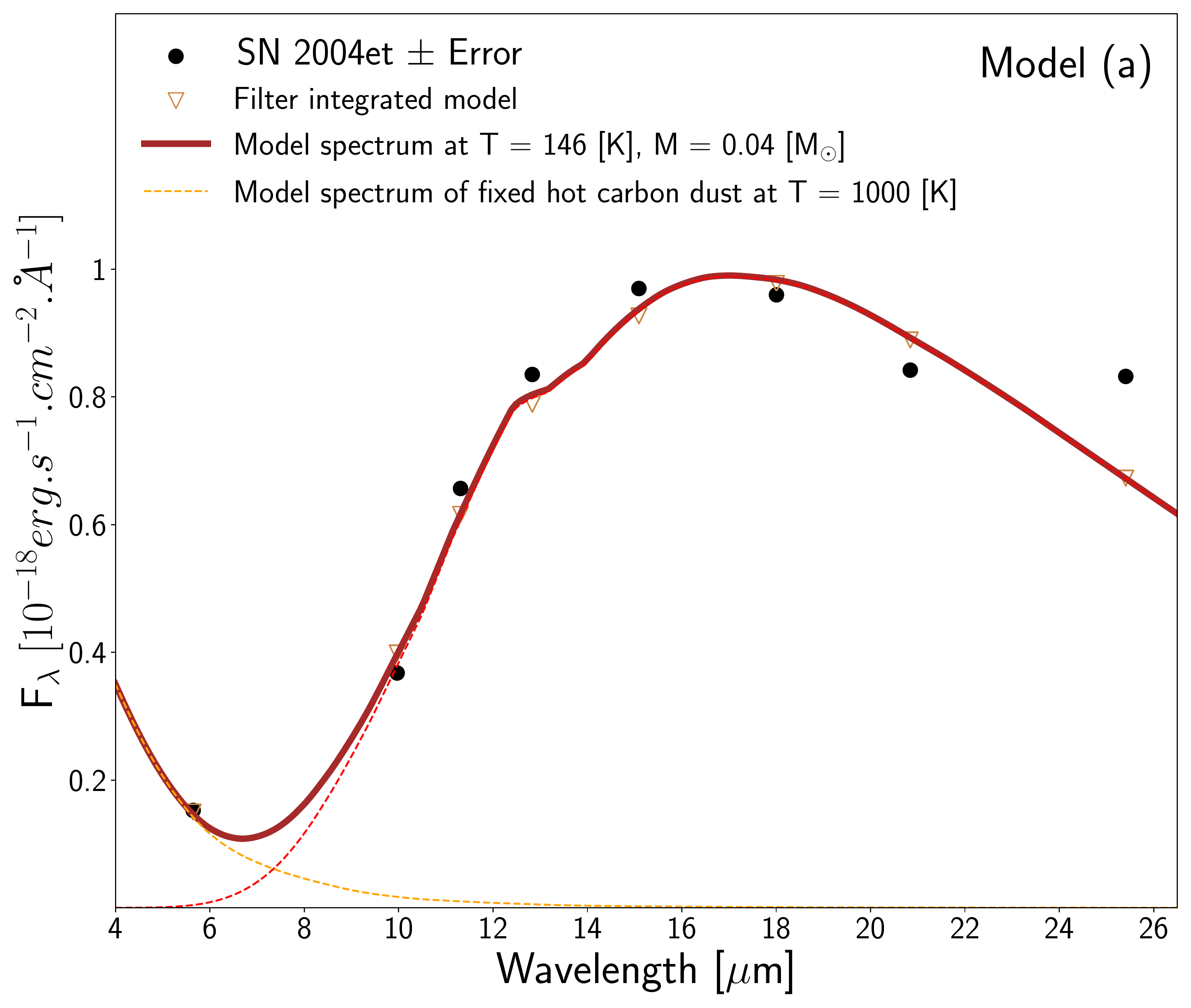}\hfill
    \includegraphics[width=0.5\textwidth]{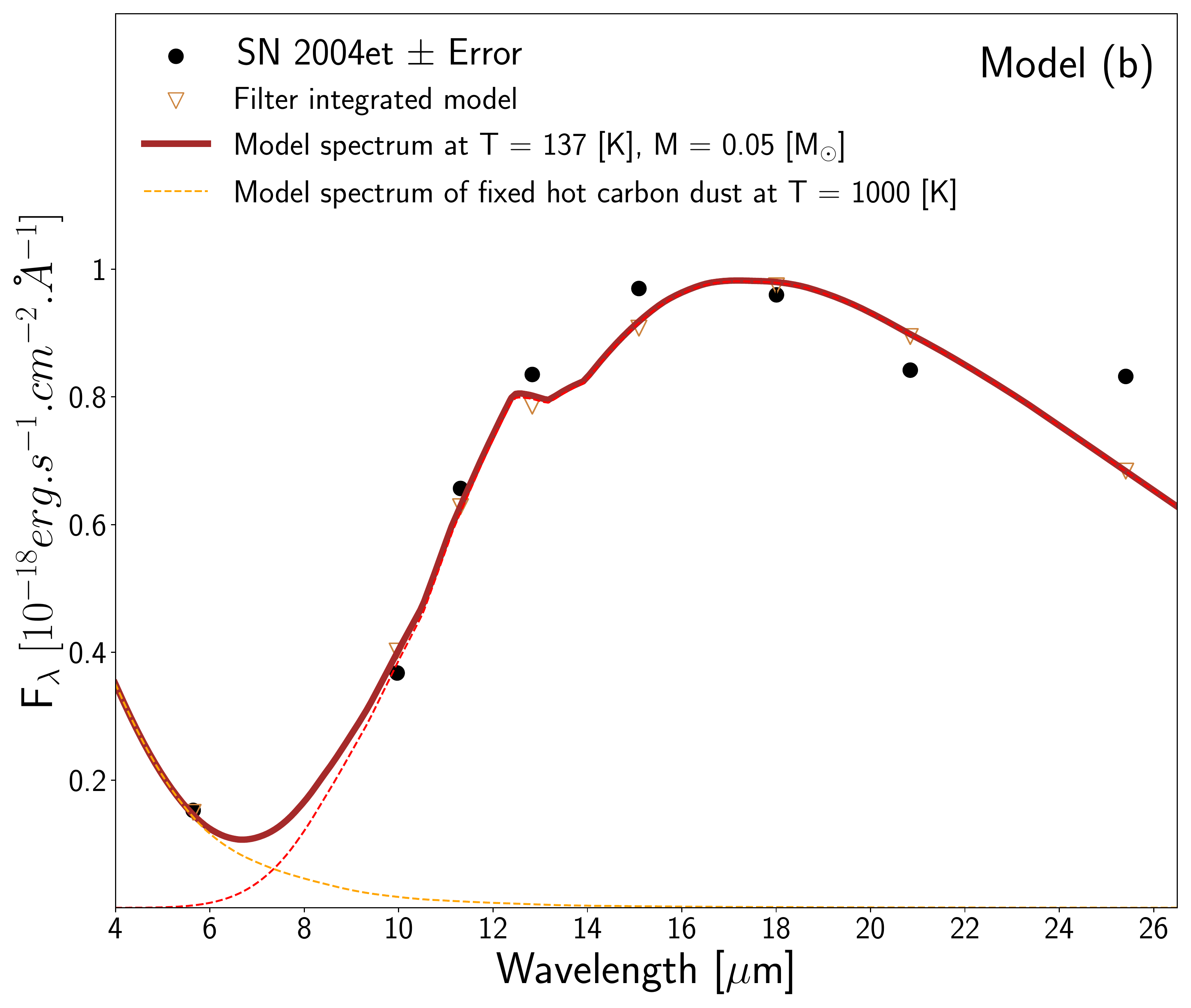}\\
        \includegraphics[width=0.5\textwidth]{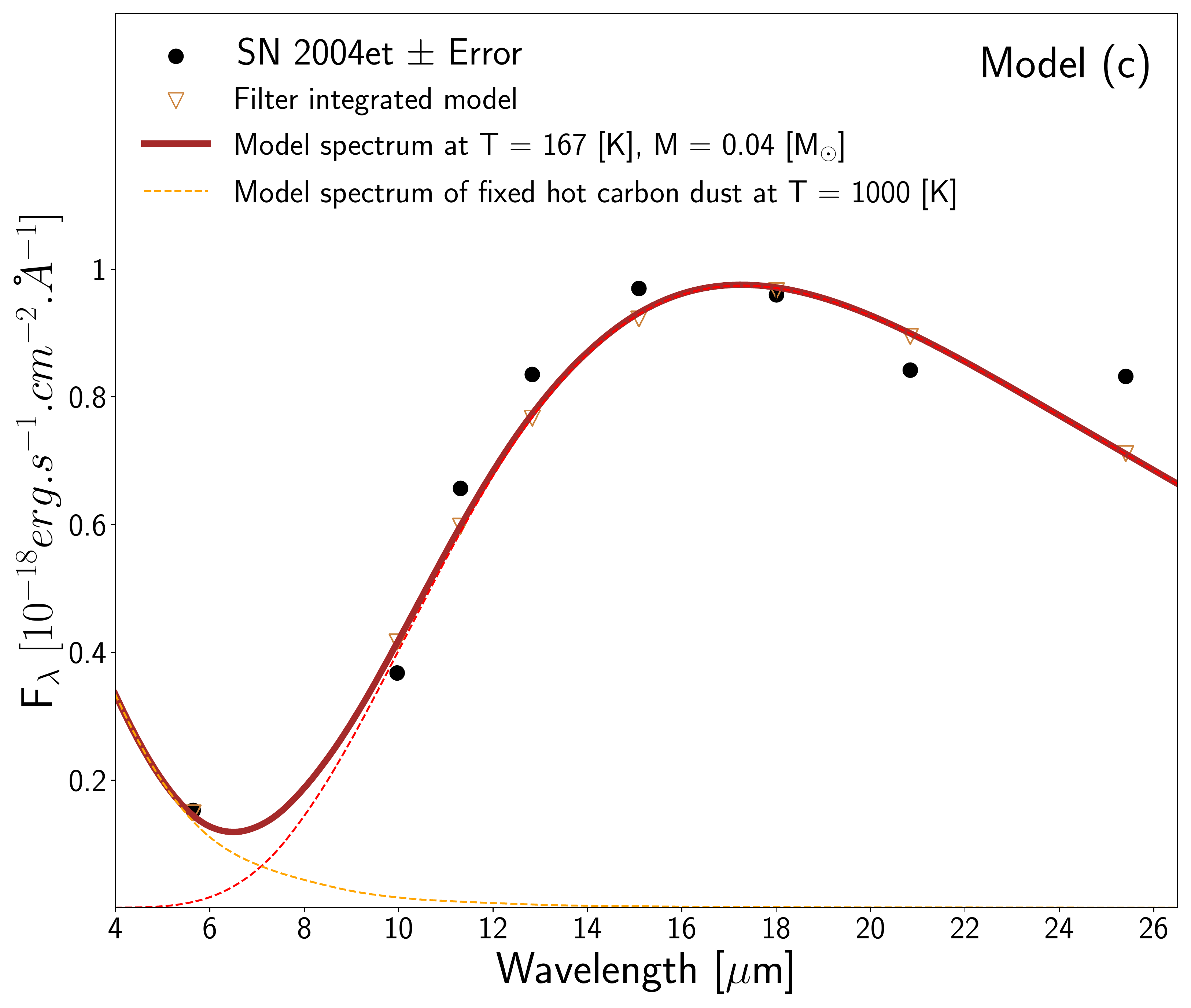}\hfill
    \includegraphics[width=0.5\textwidth]{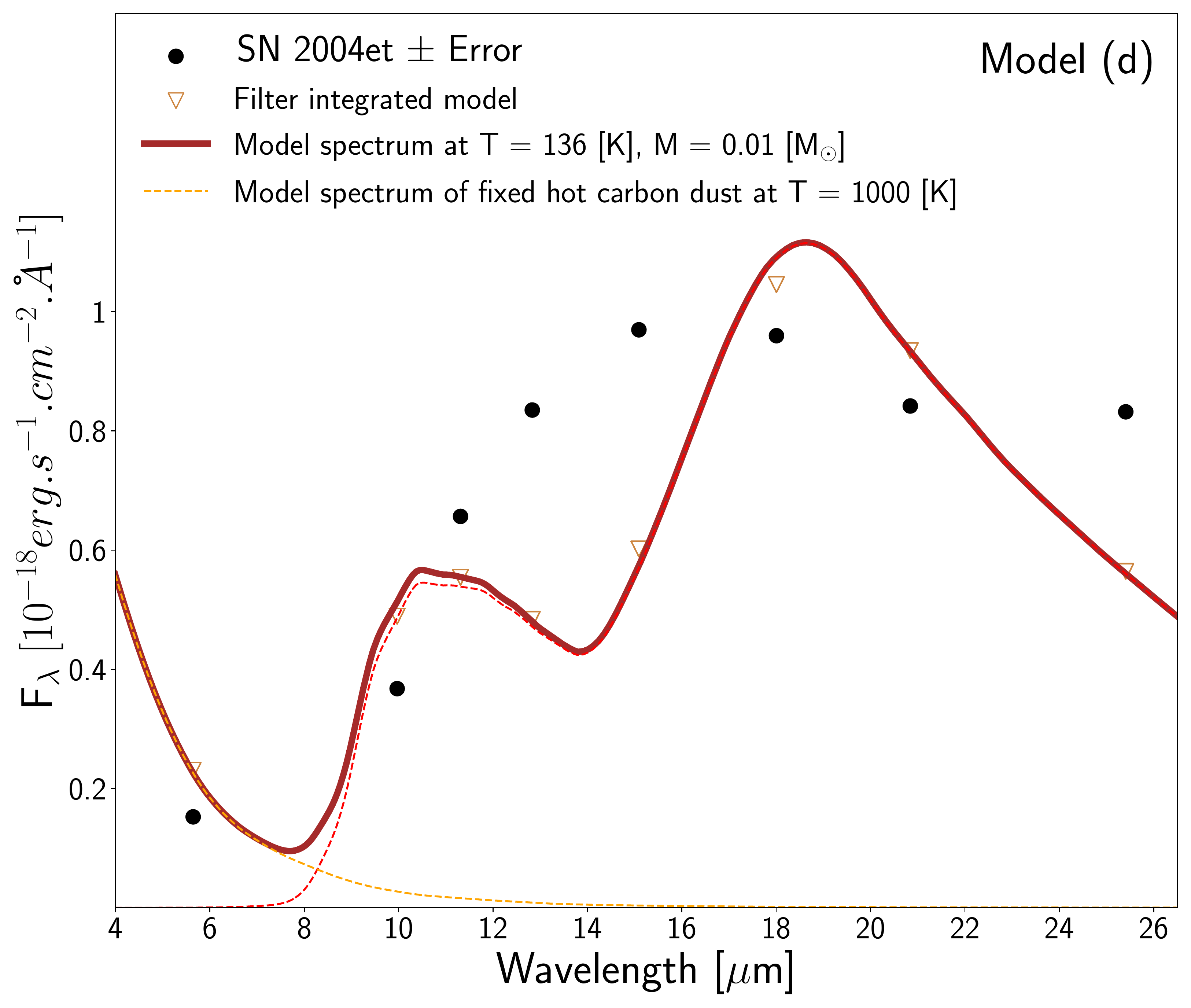}\\
    \caption{The MIR SED of SN~2004et obtained with {\it JWST}/MIRI on Sep. 20, 2022 fitted with different models/assumptions. For all four models, the solid line shows the model spectrum of amorphous C or silicates dust comprising two components. The orange dashed line shows a hot C component fixed at a temperature of 1000 K for all four scenarios, and the red dashed line shows the main C component. The resulting parameters are listed in Table~\ref{tab:model_fits04et}. \textbf{Model (a)} is assuming a dusty sphere of amorphous C using Equations~\ref{eqn:mass} and \ref{eqn:tau}, \textbf{Model (b)} shows an optically thin amorphous C dust using Equation~\ref{eqn:mass} with $P_{esc}\approx1$, \textbf{Model (c)} is assuming a dusty sphere of silicates using Equations~\ref{eqn:mass} and \ref{eqn:tau}, and  \textbf{Model (d)} is an optically thin silicates dust using Equation~\ref{eqn:mass} with $P_{esc}\approx1$.}
    \label{fig:model_fits_04et}
\end{figure*}

\begin{table*}
\centering
\caption{Best-Fitting Dust Model Parameters for SN~2017eaw
\label{tab:model_fits17eaw}}
\renewcommand{\arraystretch}{1.5}
\begin{tabular}{lcccccc}
\hline\hline
Model     &   $M_{\rm dust}$$^{\rm a}$  &   $T_{\rm hot}$$^{\rm a}$   & $T_{\rm cold}$$^{\rm a}$   &   R   &   $\tau^{\rm b}$  &   $P_{\rm esc}^{\rm b}$   \\
    &   [\Msun] &   [K] &    [K] &   [cm]    &   &   \\
\hline
(a) Dusty Sphere (Silicates) &  0.0007$^{+0.0004}_{-0.0002}$ &   1508 $\pm$ 793 & 158$^{+5}_{-5}$     &   2.87$^{+2.92}_{-0.61} \times10^{16}$   &  0 $< \tau <$ 0.8 & 0.6 $< P_{\rm esc} <$ 1\\
(b) Optically Thin Dust (Silicates) &  0.0006$^{+0.0003}_{-0.0002}$ &   800 $\pm$ 521     &  154$^{+12}_{-12}$  & -- &   $<<1$  &  $\approx1$\\
\end{tabular}
\begin{tablenotes}
      \small
      \item $^{\rm a}$ The numbers reported here are the best fit $\pm$~the 1$\sigma$ (68\%) confidence interval as reported by \texttt{lmfit.conf\_interval}. See Table \ref{tab:final_properties} for final reported numbers.\\
      \item $^{\rm b}$ $\tau$ and $P_{\rm esc}$ values have been calculated at $\lambda \approx$ 18~\um.\\
    \end{tablenotes}
\end{table*}

\begin{figure*}
    \centering
    \includegraphics[width=0.5\textwidth]{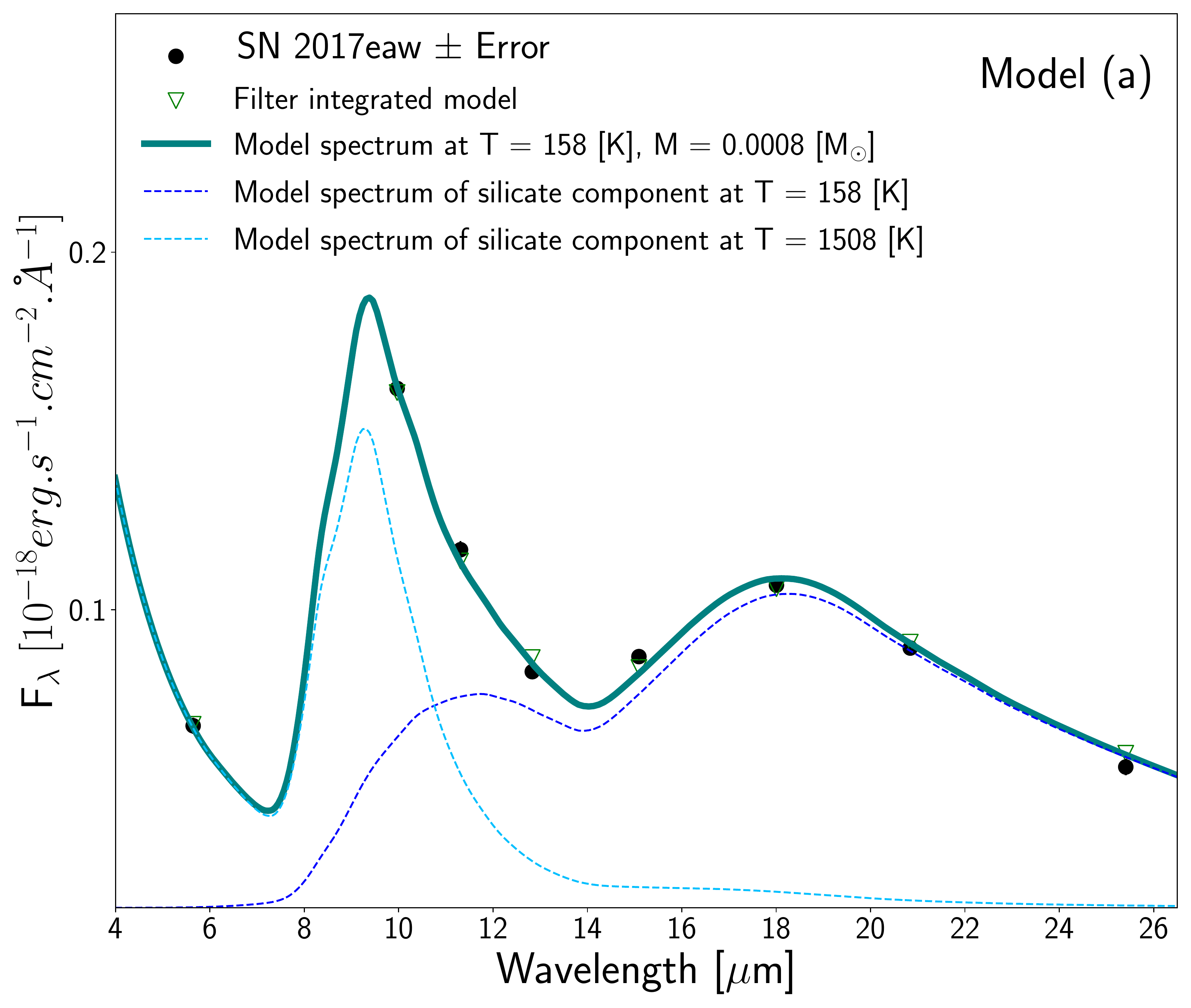}\hfill
    \includegraphics[width=0.5\textwidth]{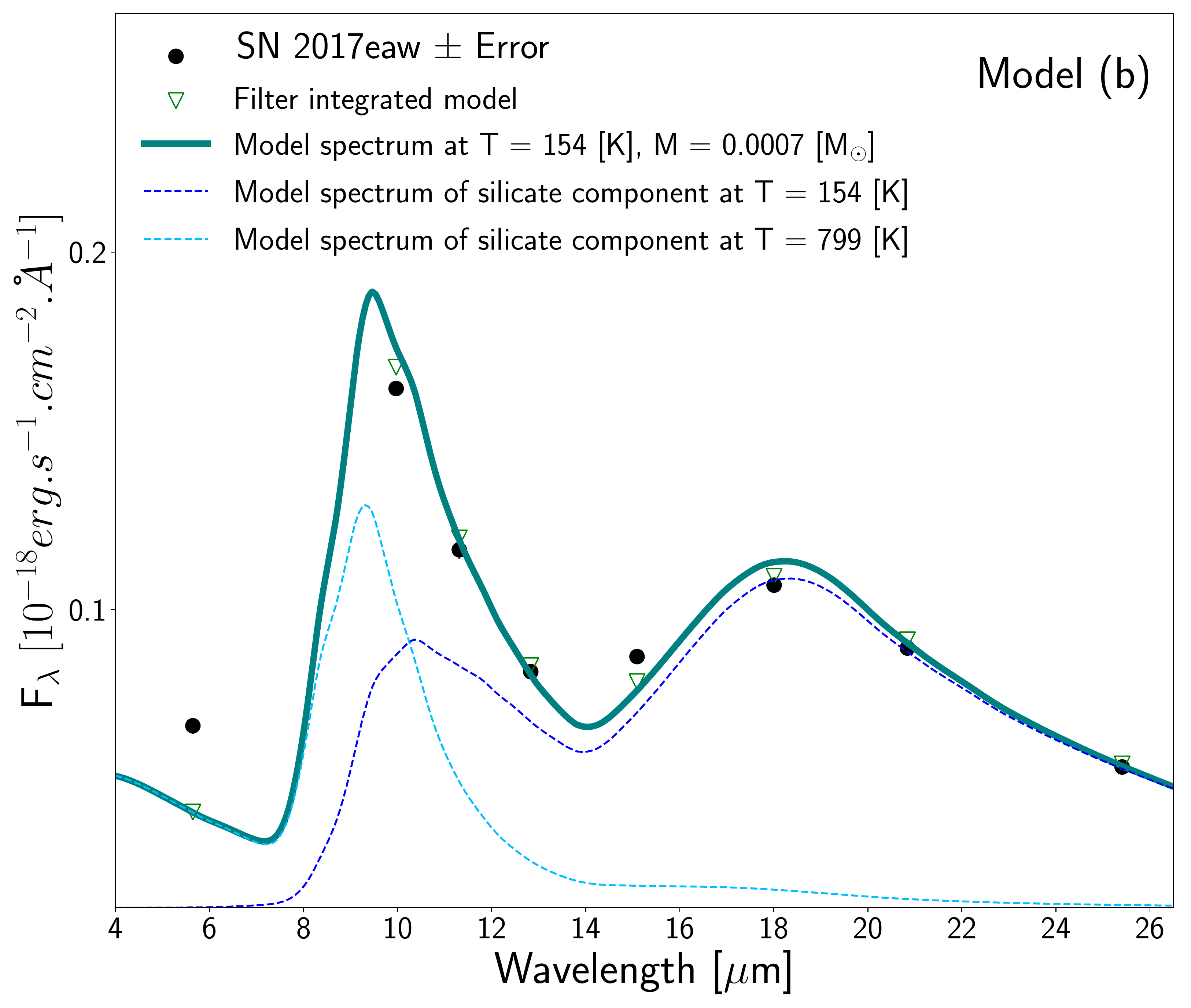}\hfill
    \caption{The MIR SED of SN~2017eaw obtained with {\it JWST}/MIRI on Sep. 21, 2022 fitted with two different models/assumptions. For both models, the solid line shows the model spectrum of silicates dust comprising two components. The dashed lines show the two components. The resulting parameters are listed in Table~\ref{tab:model_fits17eaw}. \textbf{Model (a)} is assuming a dusty sphere of silicates using Equations~\ref{eqn:mass} and \ref{eqn:tau}, and \textbf{Model (b)} shows an optically thin silicate dust using Equation~\ref{eqn:mass} with $P_{esc}\approx1$.}
    \label{fig:model_fits_17eaw}
\end{figure*}

\subsection{Fitting the Data}

To fit the IR data, we need to make some estimation about the dust composition. Since we do not have a mid-IR spectrum to base our assumptions, our choice of the chemical nature of dust is unconstrained. The many different chemical forms of cosmic dust relevant for SNe and their optical properties were explored by \cite{Arendt_2014}. In a simplified picture, observations and models of dust formation in SN ejecta, both support the presence of O-rich dust species in the form of Mg-silicates, and C-rich dust species such as amorphous carbon or graphite \citep{Wesson_2021, Ercolano_2007, Sarangi_2013}. In this study, we have also chosen Mg-silicates and/or amorphous carbon as our dust composition. The abosorption and emission properties of these grains are obtained from \cite{Draine_2007} and \cite{Zubko_2004} for silicates and amorphous carbon respectively (see \cite{Sarangi_2022} for the values of absorption coefficients $\kappa$). Given that we are dealing with mid-IR data, the wavelengths are much larger than the expected sizes of grains; in this Rayleigh regime, grain sizes do not impact the emerging fluxes. We have chosen the absorption coefficients corresponding to a grain radius of 0.1 $\mu$m. 

Using Equation \ref{eqn:thick}, we simultaneously fit the dust mass, temperature, composition, and opacity. It is important to note that the best fitting parameters are not necessarily the values that we report. Traditionally, the extragalactic SN dust community has used some variation of Equation~\ref{eqn:thick} to fit the mid-IR dust component straightforwardly with least squares minimization (i.e., using a package like Python's \texttt{lmfit}). This technique, however, can be misleading. For the case of an increasing optical depth, a significant degeneracy begins to develop. As outlined by \citet{Dwek_2019}, an almost infinite dust mass can fit the data because the emission from the dust simply does not escape. The fits quickly become unbounded. Furthermore, we have multiple dust components to our model. It is therefore not reasonable to estimate the error from the covariance matrix. Instead, we explicitly calculate the confidence intervals for variable parameters using \texttt{lmfit.conf\_interval} \footnote{https://lmfit.github.io/lmfit-py/confidence.html}. Figures \ref{fig:model_fits_04et} and \ref{fig:model_fits_17eaw} show the best fitting results. Tables \ref{tab:model_fits04et} and \ref{tab:model_fits17eaw} report the best fitting parameters, while Table \ref{tab:final_properties} reports our final dust masses as the values given by the 90\% confidence interval.

In fitting the model we assumed that all the dust radiates at the same temperature, justifying the use of Equation \ref{eqn:Pesc} for $P_{\rm{esc}}$. A more realistic scenario would be to adopt several analytical forms for the dust temperature gradient in the ejecta and use a simple radiative transfer model, such as the one used by \citet{dwek20} in modeling the emission from the galaxy Arp 220.


At first glance, SNe 2004et and 2017eaw appear to have different compositions, with SN 2017eaw having a strong 10~\micron~silicate feature. This may be true, but it may also be due to optical depth effects. Notice that an optically thick silicate dust sphere also fits the SN 2004et data quite well given the 10 \micron~feature is suppressed. From the photometry alone, it is too difficult to distinguish between models. We therefore report results for both compositions for SN 2004et.

\begin{figure}
    \centering
\includegraphics[width=0.49\textwidth]{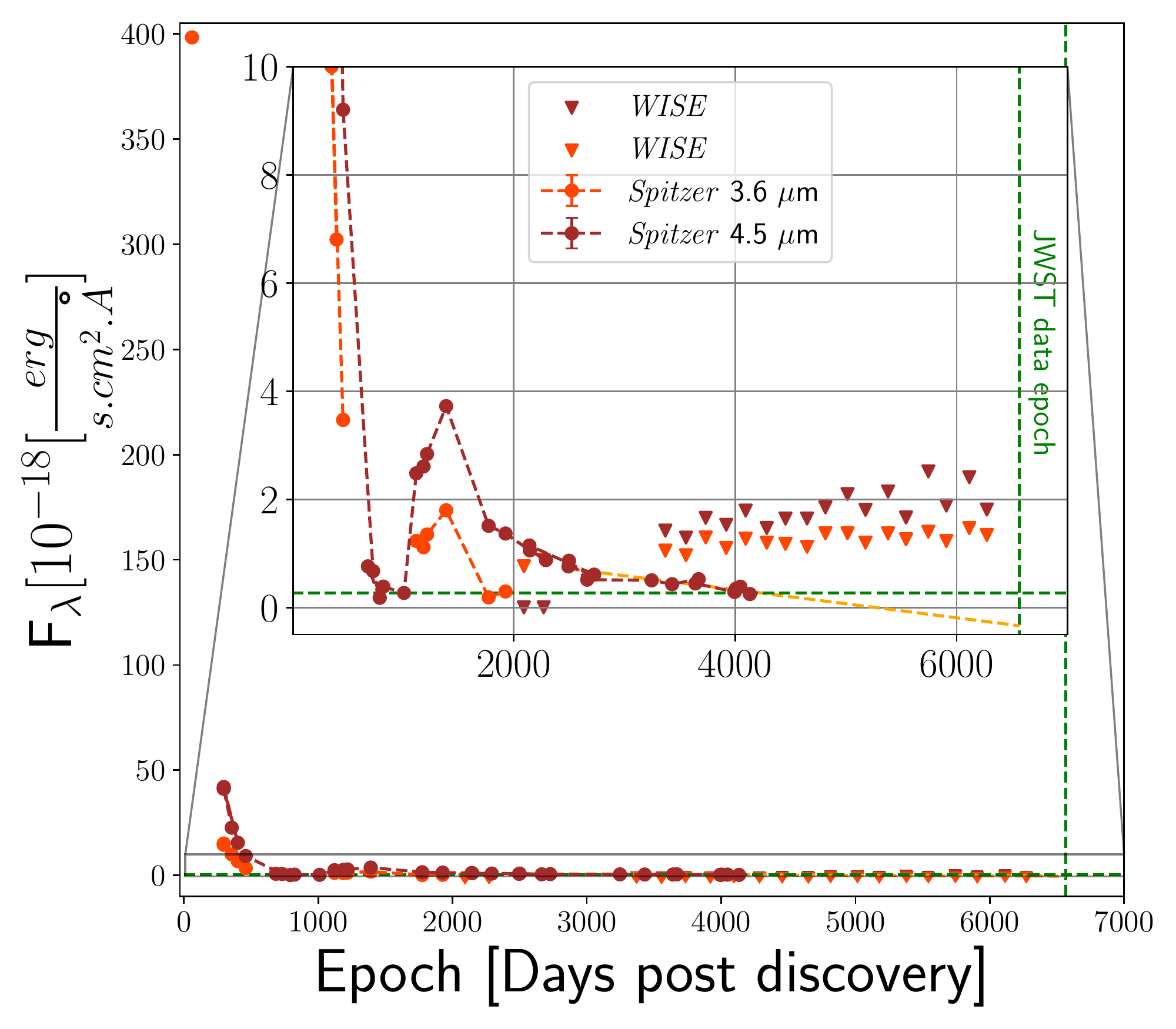}\hfill
    \caption{{\it Spitzer} and {\it WISE} photometry of SN~2004et. These data were all obtained at much earlier epochs, but help serve as a useful constraint on the presence of a hot component we use in our SN 2004et models in Figure \ref{fig:model_fits_04et}.}
    \label{fig:spitzer}
\end{figure}

For both SN 2004et and SN 2017eaw, the F560W filter showed an excess relative to the best fitting model. In both cases, we therefore add an additional, hotter contribution that we attribute to the likely presence of hot dust, either newly formed or continuously heated by ongoing shock interaction. Given the contributions of this hot component in only a single {\it JWST} filter, we have relatively few constraints. For SN 2004et, the only other long-wavelength (i.e., mid-IR) data we have are the {\it Spitzer} and {\it WISE} archival data (Sections \ref{sec:spitzer} and \ref{sec:wise}), shown in Figure \ref{fig:spitzer}.~{\it Spitzer} observations are available only through $\sim 4000$ days, while all {\it WISE}~data at epochs $>4000$ days are upper limits. The trend in the {\it Spitzer} data is unclear. Although the light curve may be steadily declining, it may also be on a plateau, which is not uncommon for SNe that have signatures of CSM interaction, although the {\it Spitzer} fluxes for SN 2004et were close to the instrumental detection limits \citep[see, e.g.][]{Fox_2011, Fox_2013, Szalai_2019, Szalai_2021}. Figure \ref{fig:spitzer} shows the possible fluxes at the epoch of the {\it JWST} observations for either scenario. We assume a flux in the range 10--18~$\mu$Jy as reasonable, which is consistent with the F560W MIRI observations. Given the limited constraints we have on this warm dust, we add a 1000~K graphite component to both models.

In addition to the warm component in SN 2004et, we see an additional rise in the F2550W data point above the expected dust model fit. We attribute this flux to the presence of yet another dust component, this time much colder. Such a cold dust component is not unexpected, given that the majority of dust is expected to cool to temperatures $<40$~K, as it did in SN~1987A \citep{Matsuura_2011}. Again, given the lack of data, we have few constraints. However, we vary the mass and temperature of this additional component and show that significant quantities of dust may be present at these longer wavelengths.

\section{Discussion}
\label{sec:discussion}

\subsection{Dust Mass Lower Limits} 
\label{sec:dust_mass}

It is worth repeating that the dust masses reported here are lower limits. We cannot account for dust hidden by high optical depths or cold dust emitting at wavelengths $>$25 \micron. Even when taking into account these lower limits, the dust mass observed in SN 2004et is still the second-largest measured dust mass based on mid-IR observations reported in any extragalactic SN to date. And aside from SN 1987A, the detections of dust in both SNe also correspond to the latest detections thus far of dust in any extragalactic SNe.

\begin{figure}
    \centering
\includegraphics[width=0.5\textwidth]{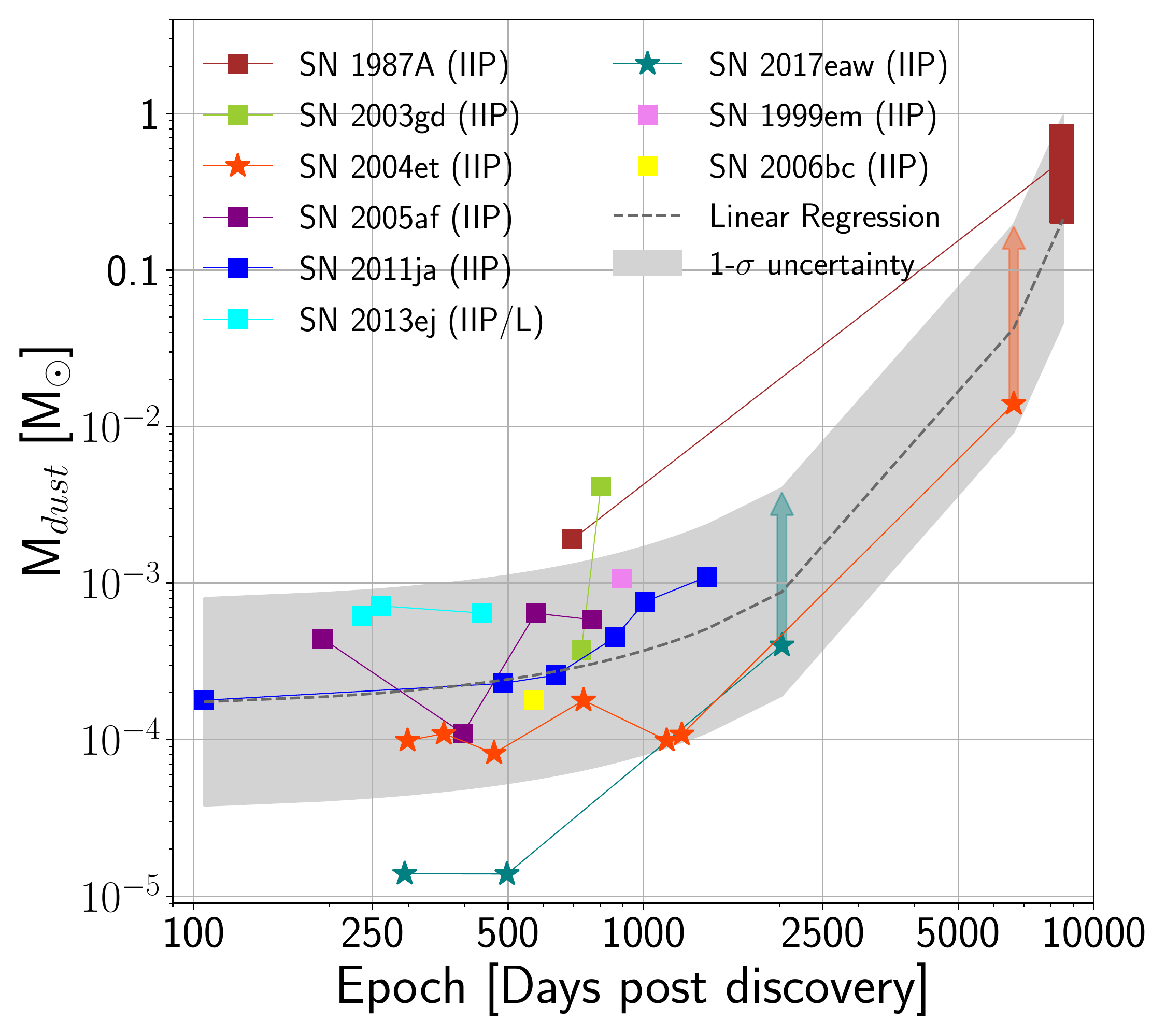}\hfill
    \caption{The dust mass in SN, as inferred from observations using the optically thin approximation, as a function of the epoch of observations. The trend may either suggest the growth of dust mass, or the increasing transparency of the ejecta with time. The vertical arrows for both SNe~2004et and 2017eaw show the possible dust reservoirs in these SNe based on our SED fits.}
    \label{fig:dustmass}
\end{figure}

It is tempting to conclude an apparent trend of increasing dust mass with time in Figure \ref{fig:dustmass}, but two important caveats should be noted. First, the dust origin and even heating mechanism is not straight-forward. Section \ref{sec:dust_geometry} explores the different possibilities. Second, the dust masses derived in previous studies are not consistent. The datasets range in mid-IR wavelength coverage and sensitivity, and the analysis techniques vary in scope, particularly how they report the best-fitting parameters versus the lower limits derived from the confidence intervals. Previous SEDs should be remodeled with a single, consistent procedure, but that work is beyond the scope of this paper. 

\subsection{Dust Origin and Heating Mechanisms} 
\label{sec:dust_geometry}

The origin and intensity of the IR emission depend on the source that heats the dust, and the geometry of the dust distribution. As noted in Section \ref{sec:scenarios}, the dust may be pre-existing in the unshocked CSM, or newly formed either in the ejecta or in the swept up shell of gas behind the SN blastwave. Assuming the distribution to be isotropic, we can impart necessary constraints on the geometry that are consistent with the physical scenarios introduced in Section \ref{sec:scenarios}. Assuming the dust to be confined within a dusty sphere, the blackbody radius, given by Equation \ref{equ:rad}, provides the smallest possible radius of the observed dust. For SN~2004et and SN~2017eaw, Equation \ref{equ:rad} yields a blackbody radius of $\sim5.4 \times 10^{16}$~cm and $1.6 \times 10^{16}$~cm, respectively.


The dust formation zone within the metal-rich ejecta is expected to be confined within 5000~km~s$^{-1}$ \citep{Truelove_1999, Maguire_2012, Sarangi_2022}, which is consistent with the velocity profile in Figure \ref{fig:opticalspec}. These velocities correspond to a radius of $\sim 2.8 \times 10^{17}$~cm and $7.8 \times 10^{16}$~cm for SN~2004et and SN~2017eaw, respectively. The ejecta radii are much larger than the blackbody radii; hence, the metal-rich ejecta is a completely acceptable site for dust that is responsible for the IR emission.

In the case of SN~2004et, \citet{Kotak_2009} and \citet{Fabbri_2011} have already done a detailed analysis on earlier optical and IR data that show increasing extinction over time that is consistent with new dust formation in the ejecta within the first 1000 days. Along these same lines, the optical spectra in Figure \ref{fig:opticalspec} show blueshifted H$\alpha$~lines that are also consistent with dust present in the ejecta, that can preferentially absorb emission from the receding shock on the far side of the SN \citep{Duvaz_2022}. In a spherical geometry, pre-existing dust, which is present beyond the blastwave radius cannot lead to this blueshift in emission lines; therefore, we argue that the dust responsible for the IR emission is most likely newly formed. 

In SNe~IIP, which are characterized by red supergiant type pre-explosion mass-loss (typically of the order of 10$^{-6}$ \Msun yr$^{-1}$), the swept up mass of gas by the SN blastwave in its first couple of decade after explosion should not exceed 0.1~\Msun. This circumstellar gas is typically H-rich, with refractory elements only amounting to about 1\%. The CSM gas behind the shock is not likely to be the site that can produce the relatively large dust masses given in Table \ref{tab:model_fits04et} and \ref{tab:model_fits17eaw}.

 Based on the these arguments, the geometry favors the dust to be present in the ejecta. However, the time frame of the dust formation within the ejecta remains unclear. With only a limited number of data points, it is not possible to differentiate between a steady, continuous formation \citep[e.g.][]{Wesson_2021}, and a relatively rapid formation followed by years of increasing optical depth as the ejecta expand \citep{Dwek_2019}. Higher cadence observations of future dust-forming SNe should be able to disentangle the two models. 

To help constrain the scenario in either case, we must also identify the heating mechanism. In the first couple of years after the explosion, newly formed dust in the ejecta is understood to be heated by the diffuse energy deposited in the ejecta, owing to the decay of radioactive $^{56}$Co. After 1500 days, the radioactive energy is dominated by the decay of $^{44}$Ti and possibly $^{57}$Co and $^{60}$Co; however, their total luminosity is expected to not exceed $10^3$~\lsolar\ \citep{Seitenzahl_2014}. In this study, we find that the mid-IR fluxes of SN~2004et and SN~2017eaw correspond to the total luminosity of  $\sim 2.2 \times 10^5$~\lsolar\ and $3.2 \times 10^4$~\lsolar, respectively. Therefore, there must be some additional source of energy that is heating the dust.


\begin{table}
\centering
\caption{Reported Dust Properties \label{tab:final_properties}}
\setlength{\tabcolsep}{3.pt}
\begin{tabular}{lccccc}
\hline\hline
SN   &   $M_{\rm dust}$$^{\rm a}$   &   R$_{\rm BB}$  &   R$_{\rm shock}$  & L$_{\rm IR}$ &  L$_{\rm opt}$  \\
     &       [\msolar]   &      [cm]  &   [cm]     & [\lsolar] &  [\lsolar]  \\
\hline
SN~2004et (C) &  $>$0.033     &   5.4$\times10^{16}$   & 2.8$\times10^{17}$ & 2.2$\times10^{5}$  & 1.4$\times10^{4}$    \\
SN~2004et (Sil) &  $>$0.014     &   5.4$\times10^{16}$   & 2.8$\times10^{17}$ & 2.2$\times10^{5}$  & 1.4$\times10^{4}$    \\
SN~2017eaw (Sil) &  $>$4$\times$10$^{-4}$     &  1.6$\times10^{16}$  &  7.8$\times10^{16}$  & 3.2$\times10^{4}$  &  1.3$\times10^{4}$  \\
\end{tabular}
\begin{tablenotes}
\small
    \item $^{\rm a}$ With 90\% confidence using \texttt{lmfit.conf\_interval}.
\end{tablenotes}
\end{table}

One source of heating is radiative emission from shock interaction. At sufficiently late times, interaction with a steady-state wind will always win against the exponential decline of radioactive decay power. Simulations by \citet{Dessart_2022} finds that a wind mass-loss rate of even $10^{-6}$~\ml\ generates a constant shock power of about $10^{40}$\,erg\,s$^{-1}$ in a standard SN~II. A fraction of that power will come out as X-rays. The thermalized part of this power will emerge primarily in the UV, especially in Ly$\alpha$ 1215.67~\AA\ and Mg\,{\sc ii}\,2800~\AA, while a few percent of that thermalized flux emerges in the optical as a weak continuum source together with emission lines, in particular H$\alpha$.

We know that both SNe 2004et and 2017eaw are undergoing late-time CSM interaction. Figure \ref{fig:opticalspec} plots the most recent optical spectra. The integrated flux for the late-time optical spectra of SNe 2004et and 2017eaw corresponds to only about $\sim10^4$~\lsolar~(Figure \ref{tab:final_properties}). The observed optical luminosity is not sufficient to heat the dust, but it does not represent the total luminosity produced by the shock-CSM interaction. \citet{Dessart_2022} shows that the shock power introduced at the interface between ejecta and CSM emerges primarily in the UV, channeled primarily into Ly$\alpha$, for which we have no observational coverage, even in the archival {\it HST}/WFC3 UVIS data. These models suggest that the observed optical luminosities ($\sim 10^4$~\lsolar) could be compatible with a UV flux of $\sim 10^5$~\lsolar, which is more than sufficient to heat the dust to the observed temperature and luminosity. 




To summarize, we propose that the most plausible scenario is that the dust is present in the ejecta, and is being heated by the interaction between the SN forward shock and the ambient CSM. Since the heating source of the dust (that is, the forward shock) is external to the ejecta dust in a spherical geometry, it is not possible to find the exact radius of the dusty sphere despite knowing the dust temperatures, as the laws of electrodynamics state that the field strength is uniform inside a sphere. The percentage of the forward shock luminosity incident on the ejecta dust depends on the ratio of velocities between the outer radius of the metal-rich ejecta and the forward shock. Owing to the very large absorption coefficient of both amorphous carbon and silicates in the UV bands \citep{Sarangi_2022}, the lower limit of the dust mass (derived in this study) is sufficient to completely absorb the incident radiation from the forward shock (can be verified by Equation \ref{eqn:Pesc}). On the contrary, the optical depths are much lower in the mid-IR bands, where we probe the SNe with {\it JWST}. Therefore, we can expect that all the absorbed radiation will be reprocessed and radiated back in the IR. For SN~2004et, using the IR luminosity of $2.9 \times 10^4$~\lsolar\ and a dust temperature of $\sim 140$~K, if we assume equilibrium of absorption and emission \citep{Temim_2013, Sarangi_2022}, the approximate forward shock velocity can be calculated to be $\sim 10,000$~km~s$^{-1}$. Comparing the IR luminosity to the forward shock luminosity (which is expected to be at least 10$^{40}$~ergs~s$^{-1}$, suggested by \citealt{Dessart_2022}), the velocity of the dusty ejecta can be found estimated as $\sim 3000$~km~s$^{-1}$, which matches well with the location of carbon-dust formation \citep{Sarangi_2022}.

Interestingly, we find that the total IR luminosity of SN~2004et is higher than the luminosity of SN~2017eaw, even though the latter is younger and is expected to have a stronger forward shock luminosity. This may indicate that the younger SN~2017eaw, being more compact, is optically thick in the mid-IR, as some theoretical studies suggest \citep{Dwek_2019,Sarangi_2022}. On the other hand, it may also indicate that the silicate-rich dust in the ejecta of SN~2017eaw is expanding at a smaller velocity (compared to carbon dust in SN~2004et), so a smaller fraction of the forward shock luminosity is absorbed here. This is again in agreement with theoretical models, which predict that silicate dust is formed in the inner regions (hence smaller velocity) of the metal-rich ejecta \citep{Sarangi_2018}.

We did not factor the impact of clumpiness of the ejecta in our analysis. The optical depths are expected to be lower in the case of clumpy ejecta \citep{Inoue_2020}. Therefore, the lower limit on the dust mass that we estimated from an optically thin scenario in this study remains completely valid. The escape probabilities of the UV photons, and therefore the heating rates, might alter when clumpiness is taken into account. We intend to address the impact of clumpiness in a future study, when more data are available to constrain our results.

Other possible scenarios exist for the heating of the dust. For example, the majority of the flux may actually come out in X-rays at late times. The reverse shock may have traveled sufficiently far back into the ejecta to be heating them directly (i.e., not radiatively). The optical depth is quite high, thereby absorbing 99\% of the optical flux. Or there may be a pulsar at the center. Each of these scenarios requires additional observations and more complex models, all of which are beyond the scope of this paper.

\section{Conclusion} \label{sec:conclusion}

In this article, we present {\it JWST} MIRI observations of SNe 2004et and 2017eaw at roughly 18~yr and 5~yr post-explosion, respectively. The mid-IR imaging unveils reservoirs of warm dust totaling $>0.014$ and $>4  \times 10^{-4}$~\msolar, respectively. When compared to other mid-IR observations of dust in SNe, these are some of the latest detections and highlight the sensitivity of {\it JWST}. Aside from SN 1987A, SN 2004et has the largest mid-IR inferred dust mass of any extragalactic SN to date. The results extend the empirical trend of an increaing dust mass in SNe~IIP at late times. Even at day $\sim 6500$, the dust in SN 2004et is still opaque ($\tau>1$), allowing for the possibility that even more dust could be detected as the ejecta continue to expand. This trend suggests that we may be able to detect $>1$~\msolar\ of dust in extragalactic SNe at $>10,000$ days, which would be sufficient to account for dust in the early universe \citep{Dwek_2007} if it can survive the reverse shock and injection back into the Universe \citep{slavin20}.

There are several caveats. First, eight filters do not offer significant constraints. For instance, the overall fit has a range of possible models (reflected in the confidence intervals in Tables \ref{tab:model_fits04et} and \ref{tab:model_fits17eaw}). The lower limit on the dust from our confidence intervals is probably a bit lower than it needs to be. The MIRI Medium Resolution Spectrograph (MRS) would help tighten these constraints and perhaps even differentiate between optically thin and thick dust models. Along these same lines, MIRI is only mostly sensitive to the warm dust. There appear to be both hotter and colder components contributing at shorter and longer wavelengths. Regardless of the confidence intervals, the dust masses reported here should therefore be considered only as lower limits. While near-IR instruments can provide shorter-wavelength data, no instruments offer longer-wavelength data. MRS, again, can offer useful constraints of the colder dust component at the longest wavelengths by providing the shape of the long-wavelength curve. 
Also, optical data are needed to examine the interaction of the shock with the circumstellar dust and gas.

The blackbody radius provides only a minimum dust radius, and we rely on this as one of our arguments for the dust originating in the ejecta. Of course, the dust shell may be much larger, asymmetric, or clumpy. Detailed radiative-transfer modeling beyond the scope of this work must take into account these possibilities to provide  consistency between the dust temperature, geometry, and heating mechanism. The heating mechanism is another unknown; in our case, the optical luminosity is insufficient to power the dust to the observed luminosity. We assume a UV component from theoretical models, but these models are untested. No SN~IIP has ever been observed at Ly$\alpha$ at late times. {\it HST} spectroscopy at these wavelengths could confirm the heating mechanism. If Ly$\alpha$ emission is not present, other possibilities must be explored, as discussed in the text.

Finally, the observed empirical trend implies significant dust growth over decades, but this trend only corresponds to the {\it inferred} dust growth. The actual dust growth may have happened at early times, with progressively more dust visible as the ejecta expand and become increasingly optically thin. Testing dust formation and evolution scenarios requires continued monitoring at multiple wavelengths. In other words, more observations are required to explore this new, exciting field of dusty SNe in the era of {\it JWST}.

\section*{Acknowledgements}
This work is based on observations made with the NASA/ESA/CSA {\it James Webb Space Telescope}. The data were obtained from the Mikulski Archive for Space Telescopes at the Space Telescope Science Institute, which is operated by the Association of Universities for Research in Astronomy, Inc., under NASA contract NAS 5-03127 for {\it JWST}. These observations are associated with program \#2666.
Some of the data presented herein were obtained at the W. M. Keck Observatory, which is operated as a scientific partnership among the California Institute of Technology, the University of California, and NASA; the observatory was made possible by the generous financial support of the W. M. Keck Foundation.
A major upgrade of the Kast spectrograph on the Shane 3~m telescope at Lick Observatory, led by Brad Holden, was made possible through generous gifts from the Heising-Simons Foundation, William and Marina Kast, and the University of California Observatories. Research at Lick Observatory is partially supported by a generous gift from Google. 
A.V.F.'s supernova group at U.C. Berkeley is grateful for financial assistance from the Christopher R. Redlich Fund and many individual donors.
T.S. has been supported by the J\'anos Bolyai Research Scholarship of the Hungarian Academy of Sciences, as well as by the FK134432 grant of the National Research, Development and Innovation (NRDI) Office of Hungary and the \'UNKP 22-5 New National Excellence Programs of the Ministry for Culture and Innovation from the source of the NRDI Fund, Hungary.
\section*{Data Availability}




\bibliographystyle{mnras}
\bibliography{myrefs} 



\appendix

\bsp	
\label{lastpage}
\end{document}